\documentclass{article}

\usepackage{amssymb}
\usepackage{amsmath}
\usepackage{hyperref}
\usepackage{pgfplots}
\usepackage{pgfplotstable}
\usepackage[capitalize]{cleveref}
\usepackage{accents}

\usepackage{mathtools}
\usepackage{amssymb}
\usepackage{xparse}
\makeatletter

\def\@@middle#1{\ifx#1.\else#1\fi}
\def\@setdelimsize#1#2{%
  \def\@argStar{#1}%
  \IfBooleanTF{#1}%
  {\def\delimsize{\middle}%
    \IfValueT{#2}{\PackageError{MathDelimiters}{Cannot provide both star and size}{Please specify one or the other}}}   %
  {\IfValueTF{#2}{\def\delimsize{#2}}{\def\delimsize{\@@middle}}}}

\def\xParseDeclarePairedDelimiter#1#2#3#4#5#6#7{
  \NewDocumentCommand{#1}{%
    #2}%
  {%
    \begingroup%
    \expandafter\IfValueTF\expandafter{#3}{#3}{\@setdelimsize{##1}{##2}}%
    \expandafter\IfBooleanTF\expandafter{\@argStar}%
    {\mathopen{}\mathclose\bgroup\left#4 #7 \aftergroup\egroup\right#5}%
    {\mathopen{\delimsize#4}#7\mathclose{\delimsize#5}}
    \endgroup%
  }%
}
\def\@space{\nonscript\,}
\newcommand\@given[1][]{%
  \@space#1\vert \allowbreak \@space\mathopen{}}
\newcommand\@suchthat[1][]{%
  \@space : \allowbreak \@space\mathopen{}}

\providecommand\given{\@space|\@space}%

\def\xParseDeclareExpectation#1#2#3#4{
  \xParseDeclarePairedDelimiter{#1}%
  {s o e{^_} m}%
  {\@setdelimsize{##1}{##2}%
    \renewcommand\given[1][\delimsize]{\@given[##1]}%
    #2\IfValueT{##3}{^{##3}}\IfValueT{##4}{_{##4}}}%
  #3#4{}%
  {\renewcommand\given{\@given[\delimsize]}%
    \@space##5\@space}}

\xParseDeclareExpectation{\E}{\mathbb{E}}[]
\xParseDeclareExpectation{\prob}{\mathbb{P}}[]
\xParseDeclareExpectation{\var}{\mathrm{Var}}[]
\xParseDeclareExpectation{\cov}{\mathrm{Cov}}[]

\xParseDeclareExpectation{\eval}{}.|

\DeclarePairedDelimiterXPP\Order[1]{\mathcal{O}}(){}{#1}
\DeclarePairedDelimiterXPP\order[1]{o}(){}{#1}
\DeclarePairedDelimiter\abs{\lvert}{\rvert}

\DeclarePairedDelimiter\p{(}{)}
\DeclarePairedDelimiter\br{\{}{\}}

\newcommand{\D}{\,{\rm d}}

\providecommand{\rset}{\mathbb{R}}

\newcommand{\nth}[1]{#1{\ensuremath{{\textnormal{'th}}}}}

\makeatother
 \usepackage{subcaption}

\usepackage{abstract} %
\usepackage{amsthm}
\usepackage[ruled]{algorithm2e}
\usepackage{algorithmic}
\usepackage{relsize}

\newcommand*{\defeq}{\stackrel{\mathsmaller{\mathsf{def}}}{=}}

\def\email#1{\href{mailto:#1}{#1}}

\definecolor{clr_no_cv}{gray}{0}
\definecolor{clr_full}{gray}{0}
\definecolor{clr_non_adaptive}{gray}{0}

\definecolor{clr_approx}{gray}{0}
\definecolor{clr_no_subsampling}{gray}{0}

\usetikzlibrary{matrix}

\pgfplotsset{
  compat=newest,
  fixed error bar/.style n args={3}{
    error bars/y dir=both,
    error bars/y explicit,
    table/x=#1,
    table/y=#2,
    table/y error plus=#3,
    table/y error minus expr={
      ifthenelse(
      \thisrow{#2} - \thisrow{#3} <= 0,
      \thisrow{#2} - 0.5 * \pgfkeysvalueof{/pgfplots/ymin},
      \thisrow{#3})
    },
  },
}

\def\MSE{\text{MSE}}
\def\tsigma{{\widetilde\sigma}}
\def\tg{{\widetilde g}}
\def\Loss{{\Lambda}}
\def\grad#1{\nabla_{\hskip -0.3em #1 \hskip 0.11em}}
\def\threshold{{\mathcal K}}
\def\hdelta{\widehat{\delta}}
\def\inner{\widehat{\textnormal{E}}}
\def\logtol2{\ensuremath{\abs{\log\tol}^2}}
\def\tol{\ensuremath{\varepsilon}}

\providecommand{\fpow}[2]{{{#1}/{#2}}}
\providecommand{\heavisidesymb}{{\ensuremath{ \textnormal{H}}}}
\DeclarePairedDelimiterXPP\heaviside[1]{\heavisidesymb}(){}{#1}

\providecommand{\changed}[1]{\begingroup\color{red}#1\endgroup}
\renewcommand{\changed}[1]{#1}

\newlength{\dhatheight}
\newcommand{\doublehat}[1]{%
    \settoheight{\dhatheight}{\ensuremath{\widehat{#1}}}%
    \addtolength{\dhatheight}{-0.35ex}%
    \widehat{\vphantom{\rule{1pt}{\dhatheight}}%
      \smash{\widehat{#1}}}}

\newcommand{\negratetriangle}[5]{%
    \draw [thick]
    ({#1}, {#2})
    -- ({#1+#3}, {#2})
    -- ({#1}, {#2 * 2^(#4 * #3)})
    node[inner sep=0,outer sep=0, anchor=south west, scale=0.8] at (#1, #2) {#5};
}

\newcommand{\posratetriangle}[5]{%
    \draw [thick]
    ({#1}, {#2})
    -- ({#1-#3}, {#2})
    -- ({#1}, {#2 * 2^(#4 * #3)})
    node[inner sep=0,outer sep=0, anchor=south east, scale=0.8] at (#1, #2) {#5};
  }

\newcommand{\axisonright}{\pgfplotsset{ylabel shift = -2ex, ytick pos=right,}}
\newcommand{\axisonleft}{\pgfplotsset{ytick pos=left,ylabel shift=-2ex}}

\newenvironment{keywords}{\par \textbf{Keywords:}}{.}
\newenvironment{AMS}{\par{ \bf AMS Class:}}{.}

\newtheorem{theorem}{Theorem}[section]
\newtheorem{remark}[theorem]{Remark}

\crefname{algocf}{Algorithm}{Algorithms}
\Crefname{algocf}{Algorithm}{Algorithms}

\usepackage{enumitem}
\usepackage[centering,hmarginratio=1:1,top=32mm,columnsep=20pt,marginparwidth=2.5cm]{geometry}

\SetAlFnt{\small}
\SetAlCapFnt{\small}
\SetAlCapNameFnt{\small}
\SetAlCapHSkip{0pt}
\IncMargin{-\parindent}

\crefname{algocf}{Algorithm}{Algorithms}
\Crefname{algocf}{Algorithm}{Algorithms}

\hypersetup{
    colorlinks,
    linkcolor={blue!50!black},
    citecolor={green!50!black},
    urlcolor={red!80!black}
  }

\setlist[enumerate]{leftmargin=.5in}
\setlist[itemize]{leftmargin=.5in}

\crefname{section}{Section}{Sections}
\crefname{subsection}{Subsection}{Subsections}
\crefname{equation}{}{}

\pgfplotsset{
  compat=newest,
  x grid style={lightgray},
  y grid style={lightgray},
  /pgfplots/error bars/error bar style={thick, solid, opacity=0.5},
  /pgfplots/error bars/error mark options={line width=0.9pt,mark
    size=3pt,rotate=90},
  every axis plot/.append style={thick, black},
  every axis plot/.append style={every mark/.append style={mark size=3,
      solid,fill opacity=0}},
  xticklabel style={text height=0.7em},
  yticklabel style={text width=2.5em,align=right},
  width=7cm,
  height=7cm,
  grid=both,
  tick align=outside,
  ylabel near ticks,
  yticklabel style={align=left},
  xtick pos=bottom,
}

\title{Sub-sampling and other considerations for efficient risk estimation in
  large portfolios}
\author{Michael B. Giles\thanks{University of Oxford
  (\email{mike.giles@maths.ox.ac.uk}).}
\and Abdul-Lateef Haji-Ali\thanks{Heriot-Watt University
  (\email{a.hajiali@hw.ac.uk}).}}
\date{\today\;\thanks{Submitted to the editors
    20 August 2020.}}

\begin{document}
\maketitle

\begin{abstract}
  \noindent
  Computing risk measures of a financial portfolio comprising thousands of
  derivatives is a challenging problem because (a) it involves a nested
  expectation requiring multiple evaluations of the loss of the financial
  portfolio for different risk scenarios and (b) evaluating the loss of the
  portfolio is expensive and the cost increases with its size. In this work,
  we look at applying Multilevel Monte Carlo (MLMC) with adaptive inner
  sampling to this problem and discuss several practical considerations. In
  particular, we discuss a sub-sampling strategy whose computational
  complexity does not increase with the size of the portfolio. We also discuss
  several control variates that significantly improve the efficiency of MLMC
  in our setting.

\end{abstract}



\section{Introduction}\label{sec:intro}

Various risk measures are computed to assess the risk of a financial
portfolio. These measures include the probability of a large loss,
Value-At-Risk (VaR) and Conditional VaR (CVaR), also called expected
shortfall. Computing these risk measures on a large portfolio usually involves
two challenges: a nested expectation and a large sum. To be more precise,
consider computing the probability that the expected loss exceeds
\changed{some given
\(\threshold_{\eta} \in \rset\)}, that is, we want to compute
\begin{equation}\label{eq:risk-measure}
  \begin{aligned}
    \eta \defeq \prob{\E{\Loss \given R_{\tau}} > \threshold_{\eta}} =
    \E{\heaviside{\E{\Loss \given R_{\tau}} - \threshold_{\eta}}},
\end{aligned}
\end{equation}
where \(\E{\Loss \given R_{\tau}}\) is the risk-neutral, expected loss given some
risk scenario, \(R_{\tau}\), at some short risk horizon, \(\tau\), and
\(\heaviside{\cdot}\) is the Heaviside function. For example, \changed{when considering
market risk, the risk scenario is the values of the underlying assets at some
risk horizon,} \(\tau\), which affect the loss incurred by the portfolio at
maturity. The loss is usually an aggregate of many losses from different
\changed{financial} derivatives depending on a set of common underlying assets. That is
\begin{equation}
  \Loss \equiv \frac{1}{P} \sum_{i=1}^{P} \Loss_{i},
  \label{eq:total-loss-as-sum}
\end{equation}
where \(P\) is the total number of derivatives and \(\Loss_{i}\) is the loss
incurred by the \(\nth{i}\) derivative. The \(1/P\) factor is a normalization
factor that ensures boundedness as the number of derivatives in the portfolio,
\(P\), increases. In realistic portfolios, the derivatives are heterogeneous
in their evaluation. Some derivatives can be computed analytically, other
\changed{derivatives have to be approximated by simulating the underlying
  assets, others still depend on assets which} can only be sampled
approximately. Moreover, the nominal values of these derivatives can vary
greatly; a few derivatives might have large nominal values and thus contribute
significantly to the total loss compared to the majority of derivatives.

A straightforward method to approximate the probability of a large expected
loss is to simulate the nested expectation in \cref{eq:risk-measure} using
Monte Carlo. That is, \(M\) independent scenarios of the risk parameter,
\(R_{\tau}\), are sampled and, for each risk scenario, \(N\) independent samples
of the total loss \(\Loss\) are sampled by evaluating the sum
in~\cref{eq:total-loss-as-sum}. This method was explored by Gordy \&
Juneja~\cite{gordy:nested} who showed that the bias in the outer expectation
is related to the variance
of the estimator of the inner expectation. See also~\cite{giorgi:ml2r} for
sharper and extended analysis of their results. Hence, using \(N\) samples to
estimate each inner expectation, \(\E{\Loss_{i} \given R_{\tau}}\), the bias in
the outer estimator is \(\Order{N^{-1} P^{-1}}\). Setting \(N =
\Order{\max\p*{1, \tol^{-1} P^{-1}}}\) and \(M=\Order{\tol^{-2}}\) to achieve
a root mean-squared (RMS) error \(\tol\), and since evaluating \(\Loss\) is an
\(\Order{P}\) operation, the total computational complexity is
\(\Order{\max\p{P \tol^{-2}, \tol^{-3}}}\). Additionally, Gordy \& Juneja
propose handling heterogeneous derivatives with different nominal values or
different computational cost in the portfolio by proportionally dividing the
\(N\) samples amongst the different derivatives instead of evaluating the
sum~\cite[Section 3.4]{gordy:nested}, see also \cref{sec:mixed-subsampling}.

In a previous work~\cite{giles:adaptive}, the authors showed how to combine
Multilevel Monte Carlo (MLMC), as introduced by Giles~\cite{giles:MLMC}, with
adaptive sampling, as introduced by Broadie et.\ al.~\cite{broadie:adapt}, to
estimate quantities of the form \(\E{\heaviside{\E{X \given Y}}}\) for two
random variables \(X\) and \(Y\). Using this strategy, for \(Y \equiv R_{\tau}\) and
\(X \equiv \Loss - \threshold_{\eta}\), the probability of a large expected loss can
be estimated with a reduced computational complexity of \(\Order{\max\p*{P
    \tol^{-2}, \tol^{-2} \abs{\log\tol}^{2}}}\). This computational complexity
is an improvement compared to that of Monte Carlo but it still suffers from
the dependence on the number of derivatives, \(P\), which, as mentioned, can
be significant for large portfolios.

{The objective of this paper is two-fold: (i) to introduce random sub-sampling
  in the context of pricing derivatives or computing risk measures and (ii) to
  show how several computational strategies can be combined in a unified
  framework for efficient computation of risk measures in large financial
  portfolios}. First, in \cref{sec:subsampling} we discuss sub-sampling
strategies to handle large sums of heterogeneous terms and present a method
whose computational complexity does not depend on the number of terms in the
sum. Then, in \cref{sec:nested} we \changed{apply this method to our
  motivating problem involving a large portfolio,} discuss several variance
reduction techniques and show how to handle different computation models for
\({\E{\Loss_{i} \given R_{\tau}}}\). In \cref{sec:mlmc-adapt} we discuss how to
apply Multilevel Monte Carlo and adaptive sampling to obtain a method whose
computational complexity is \(\Order{\tol^{-2} \logtol2}\) to achieve a RMS
error \(\tol\), independently of the number of derivatives.
\changed{Finally, in} \cref{sec:numerics}, we apply our results to fictitious portfolios
with heterogeneous derivatives to illustrate the benefit of the methods that
are presented in the current work.

\def\Term{f}
\section{Random Sub-sampling}\label{sec:subsampling}
In this section, we discuss unbiased methods to estimate an expectation
involving a sum of terms \(\br{\Term_{i}}_{i=1}^{P}\), for a large, fixed
number of terms, \(P\),
\begin{equation}\label{eq:objective-generic-avg}
  \E*{ \frac{1}{P}\sum_{i=1}^{P} \Term_i}.
\end{equation}
We focus on this generic problem in the current section and
later apply the discussed strategies to approximate the inner conditional
expectation in \cref{eq:risk-measure}, for a given risk scenario \(R_{\tau}\),
and discuss how to relate the terms \(\br{\Term_{i}}_{i=1}^{P}\) to the losses
\(\br{\Loss_{i}}_{i=1}^{P}\), depending on the computational model of
\(\E{\Loss_{i} \given R_{\tau}}\). We will initially assume that the terms
\(\br{\Term_i}_{i}\) are mutually independent (or, in the case of considering
conditional expectation, conditionally independent)
and discuss the general case later.

A na\"ive Monte Carlo estimator of \cref{eq:objective-generic-avg} with \( N \geq 1 \) samples of
the sum requires a minimum budget equal to the cost to compute the sum once. The minimum budget
thus increases with the number of terms \(P\).
Instead, we use a random sub-sampler based on the observation that
\[
  \frac{1}{P} \sum_{i=1}^P \E{\Term_i} = \E*{\frac{\Term_j}{P p_j}},
\]
where \(j\) is a random integer with \(\prob{j = i} = p_i\) for
\(i \in \br{1, \ldots, P}\) and zero otherwise.  Using \(N\) samples in a
Monte Carlo estimator to estimate \(\E{\Term_j / \p*{P\,p_j}}\), the resulting estimator is then
\[
  \frac{1}{NP}\sum_{n=1}^N \Term_{j^{(n)}}^{(n)} p_{j^{(n)}}^{-1},
\]
where \(j^{(n)}\) is the \(\nth{n}\) sample of the random integer \(j\) and
\(\Term_{i}^{\p{n}}\) is the \(\nth{n}\) sample of \(\Term_{i}\). The variance
of this estimator, which is equal to the mean-square error (\MSE{}) since the
estimator is unbiased, is
\[
  \begin{aligned}
    \var*{\frac{1}{NP}\sum_{n=1}^N \Term_{j^{(n)}}^{(n)} p_{j^{(n)}}^{-1},}
    &= \frac{1}{N P^{2}} \var{\Term_j / p_j}\\
    &= \frac{1}{N P^{2}} \p*{\sum_{i=1}^P g_i^2 p_i^{-1} - \p*{ \sum_{i=1}^P
      \E{\Term_i}}^{2}},
  \end{aligned}
\]
where \(g_i^2 \defeq \E{\Term_i^2}\). On the other hand, the \changed{expected
  total} work is \( N \sum_{i=1}^P p_i W_i \) where \(W_{i}\) is the work
required to sample the term \(\Term_{i}\). \changed{Minimizing the variance of
  the estimator subject to fixed expected total work leads to the choice
  \(p_{i} \propto g_{i} / W_{i}^{1/2}\). By using an estimate of \(g_{i}\), denoted
  by \(\tg_{i}\), and imposing the constraint of the probabilities summing up
  to 1}, we set
\begin{equation} \label{eq:optimal-p} p_i \equiv \frac{\tg_i /
    W_i^{1/2}}{\sum_{j=1}^P \tg_j /W_j^{1/2}}.
\end{equation}
The work of this random sub-sampler
is
\[
  N\ {\frac{\sum_{i=1}^P \tg_i W_i^{1/2}}{\sum_{i=1}^P \tg_i / W_i^{1/2}}}.
\]
Assuming we have a total budget \(B\) to approximate \cref{eq:objective-generic-avg}, we set
\[
  N \equiv
  {B \ \frac{\sum_{i=1}^P \tg_i / W_i^{1/2}}{\sum_{i=1}^P \tg_i W_i^{1/2}}}.
\]
Here, we ignore the restriction of the number of samples, \(N\), to integers and treat it as a
real number instead. Note that rounding the number of samples up increases the total
computational cost by \(\max_{i} W_{i}\) at most. In any case, using the previous real value of
\(N\), the optimal variance can then be bounded as
\begin{equation}\label{eq:mse-rnd-sub-sampler}
  \begin{aligned}
    \frac{1}{N P^{2}} \var{\Term_j / p_j} &\leq \frac{1}{N} \p*{\frac{1}{P} \sum_{i =1}^P \frac{g_i^2}{\tg_i}
                                            W_i^{1/2}} \p*{\frac{1}{P} \sum_{i =1}^P \frac{\tg_i}{W_i^{1/2}}} \\
                                          &\leq \frac{1}{B} \p*{\frac{1}{P} \sum_{i=1}^P \frac{g_i^2}{\tg_i} W_i^{1/2}} \p*{\frac{1}{P}
                                            \sum_{i=1}^P \tg_i W_i^{1/2}}.
  \end{aligned}
\end{equation}
If we further assume that \(g_{i} \leq c\, \tg_{i} \) for some constant \(c > 0\)
and that \( P^{-1} \sum_{i=1}^{P} \tg_{i} W_{i}^{1/2} \leq C \), for some \(C > 0\),
then the variance of the estimator is \(\Order{B^{-1}}\), independently of
\(P\), while the total cost of the estimator is \(B\), up to the rounding of
\(N\). Under these same conditions, the previous discussion applies even in
the limit as \(P \to \infty\). {For finite \(P\), we note that in the typical case
  when, for every \(i \in \br{1,\ldots, P}\), we have that \(g_{i}\) and the work
  estimate \(W_{i}\) do not increase with \(P\) and \(\widetilde g_{i}\) is
  bounded from below, we can simply use \(c \equiv
  \max_{i %
  } \p*{{g_{i}}\big/{\widetilde g_{i}}}\) and \(C \equiv
  \max_{i %
  } \widetilde g_{i} W_{i}^{1/2}\).

\subsection{Mixed sub-sampling}\label{sec:mixed-subsampling}
Another way to handle heterogeneous terms is to use deterministic, stratified sub-sampling.
This was explored in the current context of computing probabilities of a large loss by Gordy \&
Juneja~\cite[Section 3.4]{gordy:nested}. Applied to our setting, we write
\begin{equation}\label{eq:stratified-est}
  \E*{ \frac{1}{P} \sum_{i=1}^{P} \Term_i} \approx \sum_{i=1}^{P} \frac{1}{P N_{i}}
  \sum_{n=1}^{N_{i}} \Term_i^{\p{n}},
\end{equation}
where \(N_{i} \geq 1\) is the number of samples of the \(\nth{i}\) term.
The variance of this unbiased estimator is \( {P^{-2}} \sum_{i=1}^{P}
\frac{\sigma_{i}^{2}}{N_{i}}, \) where \(\sigma_{i}^{2} \defeq \var{\Term_{i}}\), while the
work is \( \sum_{i=1}^{P} N_{i} W_{i}. \) Similar to random sub-sampling, we
minimize the variance subject to a budget constraint, \(B\), to find the
optimal number of samples for the \(\nth{i}\) term
\begin{equation}
  N_{i} \equiv %
  {B \cdot \frac{\tsigma_{i} / W_{i}^{1/2}}{ \sum_{j=1}^{P} \tsigma_{j}\, W_{j}^{1/2}}},\label{eq:stratified-optimal-N}
\end{equation}
assuming we have estimates of \(\sigma_{i}\) denoted by \(\tsigma_{i}\). Note that
we again ignore the integer constraints on \(N_{i}\) and treat it as a real
number. The optimal variance is bounded by
\begin{equation}\label{eq:MSE-stratified}
  \frac{1}{B} \p*{\frac{1}{P}\sum_{i=1}^P \frac{\sigma_i^2}{\tsigma_i} W_i^{1/2}} \p*{\frac{1}{P} \sum_{i=1}^P \tsigma_i
    W_i^{1/2}},
\end{equation}
assuming \(N_{i} \geq 1\) for all \(i\). If we further assume that \(\sigma_{i} \leq c\,
\tsigma_{i} \) for some constant \(c\) and that \( P^{-1} \sum_{i=1}^{P}
\tsigma_{i} W_{i}^{1/2} \leq C \) for some \(C > 0\), then the variance is
\(\Order{B^{-1}}\), independently of \(P\) and similar to random sampling.
However, a crucial constraint is that the budget, \(B\), must be sufficiently
large so that \(N_{i} \geq 1\) in~\cref{eq:stratified-optimal-N} for all \(i\),
\changed{otherwise the estimator~\cref{eq:stratified-est} is biased}.
In particular, the budget must be at least \(\sum_{i=1}^{P} W_{i}\) to have at
least one sample per term. This leads to a computational complexity that
depends on the number of terms in the sum, unlike random
sub-sampling. %
\changed{On the other hand, the} variance of the stratified sub-sampler in
\cref{eq:MSE-stratified} is always smaller than the variance of the random
sub-sampler in \cref{eq:mse-rnd-sub-sampler}. The variance reduction roughly
scales with \( {P^{-1}}\sum_{i=1}^{P} \p{\widetilde g_{i}-\widetilde
  \sigma_{i}}W_{i}^{1/2}\) which is bounded independently of \(P\). In other words,
in our setting, using random sub-sampling rather than stratified sub-sampling
increases the error by a constant independent of \(P\). %

\changed{We can also} combine random and stratified sub-sampling as follows
\[
  \E*{\frac{1}{P} \sum_{i=1}^{P} \Term_{i}} = \E*{\frac{1}{P} \sum_{i=1}^{K} \Term_{i}} + \frac{1}{P}
  \E{\Term_{j} / p_{j}},
\]
where \(\prob{j=i} = p_{i}\) for \( j \in \br{K+1, \ldots, P}\) and is zero
otherwise. Then the sum of the first \(K\) terms is approximated using
stratified sub-sampling while the sum of the remaining \(\p{P-K}\) is
approximated using random sub-sampling. \changed{Compared to random
  sub-sampling, this new sub-sampler evidently leads to smaller variance for a
  fixed budget }when the \( K \) terms are themselves deterministic, i.e.,
\(\E{\Term_{i}} = \Term_{i}\) for \(i \leq K\). In this case, evaluating the sum
of the \(K\) terms directly increases the work by \(\sum_{i=1}^{K} W_{i}\) but
decreases the variance %
by \(\sum_{i=1}^{K} \tg_{i} W_{i}^{1/2}\), approximately. Assuming the budget is
larger than \(\sum_{i=1}^{K} W_{i}\) and by picking those \(K\) terms to have
large \(\tg_{i} / W_{i}^{1/2}\), i.e., large nominal value or small cost, we
can ensure the increase in cost is small compared to the decrease in the
error.
{To further illustrate this point, consider the case when \(W_{i}=1\) for all
  \(i=1,\ldots, P\) and \(\br{f_{i}}_{i=1}^{P}\) are all deterministic, i.e., we
  are simply estimating the average \( P^{-1} \sum_{i=1}^{P}f_{i} \) using a
  computational budget \(B \leq P\); when \(B \geq P\) we can compute the average
  directly.
The mixed sub-sampler can then be written as
\[
  \frac{1}{P} \sum_{i=1}^P f_i \approx \frac{1}{P} \sum_{i=1}^{K} f_{i} + \frac{P-K}{P
    \,\p{B-K}} \sum_{n=1}^{B-K} f_{j^{(n)}},
\]
for \(K \geq 0\) and where $j$ is a random integer over $\br{K+1,\ldots, P}$. The
variance is
\[
  \begin{aligned}
    \frac{\p{P-K}^{2}}{P^{2} \, \p{B-K}} \cdot \var{f_{j}}
    \leq \p{\max_{i} f_{i}^{2}} \, \frac{\p{P-K}^{2}}{P^{2}\,\p{B-K}}.
  \end{aligned}
\]
The optimal value of $K$ which minimizes the variance
is %
\(\min(0, 2B-P)\) and the corresponding variance is bounded by
\[
  \p{\max_{i} f_{i}^{2}} \cdot \begin{cases}
                             4 \, \p{P-B} \big/ {P^{2}} & P/2 \le B \leq P\\
                             1 \big / B & 0 < B\leq P/2. \\
  \end{cases}
\]
This is consistent with intuition: when the computational budget passes a
certain threshold, in this case \(P/2\), sub-sampling some terms
deterministically \changed{leads to smaller variance for the same computational
  budget}. %
More generally, %
determining if a particular term \(f_{i}\) should be sub-sampled
deterministically or randomly for a given budget \(B\) requires good estimates
of both \(\widetilde g_{i} \approx g_{i}\) and \(\widetilde \sigma_{i} \approx \sigma_{i}\) (compare
\cref{eq:mse-rnd-sub-sampler} and~\cref{eq:MSE-stratified}), and hence of
\(\E{f_{i}}\), the quantity we are trying to estimate. %
If the optimal strategy is to sub-sample \(f_{i}\) deterministically instead
of randomly, the variance reduction roughly scales with the difference,
\(\p{\widetilde g_{i} - \widetilde \sigma_{i}}W_{i}^{1/2}\). Considering the need
for additional estimates, the optimization of the sub-sampling strategy for a
term, \(f_{i}\), is worthwhile only when the budget is sufficiently large
compared to the number of term \(P\)
and we know that \(\p{\widetilde g_{i}-\widetilde \sigma_{i}}W_{i}^{1/2}\) is
large, which is maximal when \(f_{i}\) is deterministic.
Hence, when considering a portfolio of terms, the variance reduction will be
significant if the portfolio contains mostly deterministic terms or terms with
small variability. Additionally, using mixed sub-sampling complicates analysis
and precludes the application of other computational methods, such as using
antithetic sub-sampling in MLMC, c.f. \cref{sec:mlmc-adapt}.
Based on these observation, and several numerical experiments, we have found
that mixed sub-sampling is not worthwhile in most practical cases, including
the example that we consider in \cref{sec:numerics}. }

\subsection{Dependent \texorpdfstring{\(\Term_{i}\)}{f}}
In the beginning of this section, we assumed that \(\br{\Term_{i}}_{i=1}^{P}\) are mutually
independent. In real applications, including the ones we consider in this work, some of these
terms might depend on a set of common underlying random variables. Nevertheless, we can use
independent samples of those underlying random variables when sampling \(\Term_{i}\) to get
independent samples of \(\Term_{i}\) and the previous discussion applies. Clearly such
re-sampling introduces additional overhead since we have to re-sample the common underlying
random variables.

On the other hand, this re-sampling has several advantages. In addition to simplifying analysis
and implementation and making the parallelization of the sampler easier, Gordy \&
Juneja~\cite[Section 3]{gordy:nested} argue that re-sampling the common random factors is
advisable to ensure that the Monte Carlo errors cancel out at the portfolio level.
Another advantage is that this re-sampling allows us to optimize the number of
samples \emph{per term} based on estimates of the second moments or variance
of \(\br{\Term_{i}}_{i=1}^{P}\). Because of these advantages,
we argue %
that re-sampling is the prudent choice in most situations. It should be noted however that
terms that are known to be negatively correlated should be sampled together to reduce the
overall variance and hence the computational cost. In \cref{sec:nested} we will see
additional strategies to reduce the variability of the loss variables, \(\Loss_{i}\), in
certain settings.
\def\Term{f}
\def\payoff{h}
\section{Probability of Loss as a Nested Expectation}\label{sec:nested}
In this section, we focus on our motivating problem of evaluating the
probability of a large loss of a financial portfolio under market risk. We
will focus on a model for the loss of a derivative that can be written as a
\changed{difference between \(V_{i, \tau}\), the discounted value of the
derivative given the risk scenario, \(R_{\tau}\), at the risk horizon, \(\tau\), and
\(V_{i,0}\), the risk-neutral discounted value at initial time. That is}
\[
  \begin{aligned}
    \changed{\E_{\mathbb Q}{\Loss_{i} \given R_{\tau}}} &= V_{i, 0} - V_{i, \tau}\\
    &= \E_{\mathbb Q}{ \payoff_{i}\p{S}} - \E_{\mathbb Q}{ \payoff_{i}\p{S}
      \given S(\tau) = R_{\tau}}.
  \end{aligned}
\]
\changed{Here, \(\mathbb Q\) is the risk-neutral measure and \(\payoff_{i}\)
  is the discounted payoff functional which depends on the asset process,
  \(S\)}. %
We will also assume that \(S\) is a stochastic process satisfying an It\^o
stochastic differential equation (SDE)
\begin{equation}\label{eq:S-sde}
  \D S\p{t} = a\p{t, S\p{t}} \D t + b\p{t, S\p{t}} \D B\p{t},
\end{equation}
\changed{for some sufficiently smooth coefficients}, \(a\) and \(b\), and a
Brownian process, \(\br{B\p{t}}_{t \geq 0}\).
Recall that we are interested in computing
\[
  \begin{aligned}
    \eta &= \prob{\E_{\mathbb Q}{\Loss \given R_{\tau}} > \threshold_{\eta}} \\
      &= \E_{\mathbb P}{ \heaviside{\E_{\mathbb Q}{ \Loss-\threshold_{\eta}\given R_{\tau}}}}\\
      &= \E*_{\mathbb P}{ \heaviside*{\E*_{\mathbb Q}{ \frac{1}{P} \sum_{i=1}^{P} \Loss_{i}
        -\threshold_{\eta}\given R_{\tau}}}},
  \end{aligned}
\]%
for a given \changed{\(\threshold_{\eta} \in \rset\) and \(\mathbb Q\) and
  \(\mathbb P\) being the risk-neutral and phyical measures, respectively.
  S}ince we consider the market risk, the risk parameter, \(R_{\tau}\), is the
asset value, \(S\p{\tau}\), in the physical measure, \(\mathbb P\), at
\changed{the risk horizon} \(\tau\).

We will consider three common categories of computation models for \(\E_{\mathbb Q}{\Loss_{i} \given
  R_{\tau}}\) and, for each computation model, we will discuss different strategies to reduce the
variability of \(\Loss_{i}\) which in turn reduces the bias of a Monte Carlo estimator of
\(\eta\), as discussed in the introduction. At the end of this section, we will construct a
``portfolio of terms'', \(\br*{\Term_{i}}_{i=1}^{P}\), such that
\[
  \E*_{\mathbb Q}{\frac{1}{P} \sum_{i=1}^{P} \Loss_{i} \given R_{\tau}} = \E*_{\mathbb Q}{\frac{1}{P}
    \sum_{i=1}^{P} \Term_{i} \given R_{\tau}}.
\]
Then we can apply the sub-sampling strategies that were discussed in the
previous section when computing the inner expectation of the sum. Recall that
when using a random sub-sampler to estimate the right hand side in the
previous equation the optimal probabilities depend on estimates of the work
required to sample \(\Term_{i}\) and of \(g_{i}^{2} = \E_{\mathbb
  Q}{\Term_{i}^{2} \given R_{\tau}}\) for every \( i \in \br{1, 2, \ldots, P} \), i.e.,
estimating \(g_{i}\) ultimately depends on the risk scenario. For an estimator
of \(\eta\) which is based on sampling many risk scenarios this is clearly too
costly, with a cost that grows with \(P\) which is counter to our original
objective of devising a method whose computational complexity does not depend
on \(P\). Instead, we propose to use estimates \(\tg_{i} \approx g_{i}\) that do not
depend on the risk scenario. For example, we may assign them to values that
represent the relative importance of an derivative compared to the others, or
we may assign \( \tg_{i} = \E_{\mathbb Q}{\Term_{i}^{2}}\) for all \(i\) and
all risk scenarios.

\subsection{Exact, deterministic evaluation}\label{sec:exact-computation}
For some derivatives, \(\Loss_{i}\) might be deterministic when conditioned on
the risk scenario \(R_{\tau}\), or we may be able to directly, with unit cost,
compute \(\E_{\mathbb Q}{\Loss_{i} \given R_{\tau}}\) exactly, or almost exactly,
given the risk scenario \(R_{\tau}\). For example, when considering put or call
options on assets that follow Geometric Brownian \changed{processes}, we may be able to
solve the Black-Scholes partial differential equation (PDE) analytically or
numerically with sufficient accuracy. Note that, %
the Black-Scholes PDE needs to be solved only once to compute \(\E_{\mathbb
  Q}{\Loss_{i} \given R_{\tau}}\) for all risk scenarios \(R_{\tau}\), hence we may
consider approximating the solution to the PDE as \emph{offline work}. In this
case, we set \(\Term_{i} \equiv \E_{\mathbb Q}{\Loss_{i} \given R_{\tau}}\) for a
given \(R_{\tau}\). Note that for a given risk scenario \(R_{\tau}\), \(\Term_{i}\)
is deterministic with zero variance and the cost to compute it is
\(\Order{1}\).

\paragraph{Delta Control Variate}
Using the Delta Greek to construct a control variate for the probability of
large loss is well-known, c.f,~\cite{glasserman:monte-fin, gou:ms-var}, and we
recall the basic idea here. Recall that the expected loss incurred by
derivative \(i\) given a risk scenario, \(R_{\tau}\), is written as a difference,
i.e., \( \E_{\mathbb Q}{\Loss_{i} \given R_{\tau}} \equiv V_{i, 0} - V_{i, \tau} \).
Then, using an It\^o expansion yields
\changed{\[
  \E_{\mathbb Q}{\Loss_{i}^{2} \given R_{\tau}}
  = \p*{\p*{R_{0} - R_{\tau}} \cdot \grad{R_{0}} V_{i, 0}}^{2} + \Order{\tau^{2}},
\]}
where \changed{\(R_{0} \equiv S\p{0}\) and}, for \(R_{\tau}\) being the price of the
underlying asset, \(\nabla_{R_{0}} V_{i, 0}\) is the Delta Greek.
The first term dominates in the previous expression since the risk parameter
is an It\^o process, \(R_{\tau} \equiv S\p{\tau}\), yielding \(\E{\abs{R_{\tau} -
    R_{0}}^{2}} = \Order{\tau}\). %
By subtracting this term, we can define a new loss variable,
\changed{\(\widehat{\Loss}_{i} \defeq \Loss_{i} - \p*{R_{0} - R_{\tau}} \cdot
  \grad{R_{0}} V_{i, 0}\), for a given risk scenario,
  \(R_{\tau}\),} %
and a new loss threshold, which depends on the risk scenario,
\begin{equation}\label{eq:loss-threshold-delta}
  \begin{aligned}
    &&&\widehat{\threshold}_{\eta} \defeq \threshold_{\eta} - \p*{R_{0} - R_{\tau}} \cdot \grad{R_{0}} V_{0}\\
    \text{where} &&&\grad{R_{0}} V_{0} = \frac{1}{P} \sum_{i=1}^{P} \grad{R_{0}} V_{i, 0}.
  \end{aligned}
\end{equation}
So that
\[
  \E*_{\mathbb Q}{\frac{1}{P}\sum_{i=1}^{P} \widehat{\Loss}_{i} -
    \widehat{\threshold}_{\eta} \given R_{\tau}} = {\E*_{\mathbb
      Q}{\frac{1}{P}\sum_{i=1}^{P} \Loss_{i} - \threshold_{\eta}\given R_{\tau} }}, \]
with \(\E_{\mathbb Q}{\widehat{\Loss}_{i}^{2} \given R_{\tau}} = \Order{\tau^{2}}\).
Hence, we have the deterministic term \({\Term_{i} \equiv \E_{\mathbb
    Q}{\widehat{\Loss}_{i} \given R_{\tau}}}\) with a second moment
\(\Order{\tau^{2}} \ll \Order{\tau}\) since \(\tau \ll 1\).
Note that \(\grad{R_{0}} V_{i, 0}\) is independent of the risk scenario,
\(R_{\tau}\), for all \(i\) and can be computed once for all risk scenarios as
\emph{offline work}. If the portfolio is delta-hedged then \(\grad{R_{0}}
V_{0} =0\).

\subsection{Exact simulation}\label{sec:exact-sim}
In some settings, we might be able to exactly sample \(\Loss_{i}\) for a given
risk scenario \(R_{\tau}\), but cannot compute \(\E_{\mathbb Q}{\Loss_{i} \given
  R_{\tau}}\) exactly. This is the case for example for exotic options or
underlying assets involving high dimensional It\^o processes, but when we
might still be able to solve the underlying SDEs analytically to exactly
sample \(\Loss_{i}\) for a given \(R_{\tau}\), e.g., when the SDE solution is a
Geometric Brownian Motion. In this case, we simply set \(\Term_{i} \equiv
\Loss_{i}\). Note that, for a given risk scenario \(R_{\tau}\), the term
\(\Term_{i}\) has non-zero variance and the cost to compute it is again
\(\Order{1}\).

\paragraph{Reducing the variance of
  \texorpdfstring{\(\Loss_{i}\)}{...}}\label{sec:loss-sample-cv}
\changed{Denote by \(S_{t, x}\) the solution of~\cref{eq:S-sde} given \(S\p{t} = x\),
then we can write}
\[
  \begin{aligned}
    \Loss_{i} &= \payoff_{i}\p{S_{0, R_{0}}} - \payoff_{i}\p{S_{\tau,R_{\tau}}}\\
    &= \payoff_{i}\p*{S_{\tau, S \p{\tau}}} - \payoff_{i}\p*{S_{\tau,R_{\tau}}}.
\end{aligned}
\]
Hence, to sample \(\Loss_{i}\) for a given risk scenario \(R_{\tau}\), we need to
first sample \(S\p{\tau}\), which requires sampling a Brownian path
\(\br*{B\p{t}}_{0 \leq t \leq \tau}\). Then, we sample \(\br{S_{\tau, S\p{\tau}}\p{t}}_{t \geq
  \tau}\) and \(\br{S_{\tau, R_{\tau}}\p{t}}_{t \geq \tau}\) starting from \(S\p{\tau}\) and
\(R_{\tau}\), respectively, which requires sampling one shared Brownian path
\(\br*{B\p{t}}_{t \geq \tau}\). While we could use \changed{two independent Brownian
  paths to sample two independents paths \(S_{\tau, S\p{\tau}}^{\p{1}}\) and \(S_{\tau,
    R_{\tau}}^{\p{2}}\),
  this would yield a larger second moment. For example when \(\payoff_{i}\p{S} \equiv
  \payoff_{i}\p{S\p{T}}\) for some maturity, \(T \gg \tau\), i.e., the payoff is a
  function of the asset value at maturity, and for a sufficiently smooth payoff
  functional, \(\payoff_{i}\), we have
  \[
    \begin{aligned}
      \E*_{\mathbb Q}{\Loss_{i}^{2} \given R_{\tau}} &= \E*_{\mathbb
        Q}{\abs{S\p{\tau} - R_{\tau}}^{2} \given R_{\tau}} %
      + \Order*{\E*_{\mathbb Q}{\abs{S_{\tau, R_{\tau}}^{\p{1}}\p{T} - S_{\tau,
              R_{\tau}}^{\p{2}}\p{T}}^{2} \given R_{\tau}}} %
      .
    \end{aligned}
  \]
  Here, the second term dominates since \(T \gg
  \tau\). %
  Using a shared Brownian path to sample \\ \(\br{S_{\tau, S\p{\tau}}\p{t}}_{t \geq
    \tau}\) and \(\br{S_{\tau, R_{\tau}}\p{t}}_{t \geq \tau}\) and for a sufficiently smooth
  payoff functional, \(\payoff_{i}\), we write
  \[
    \begin{aligned}
      \E{\Loss_{i}^{2} \given R_{\tau}} &\leq& 2\, \E{\p{\payoff_{i}\p{S_{0, R_{0}}} -
          \payoff_{i}\p{S_{0, R_{\tau}}}}^{2} \given R_{\tau}} %
      &\ +\ && %
      2\, \E{\p{\payoff_{i}\p{S_{0, R_{\tau}}} - \payoff_{i}\p{S_{\tau, R_{\tau}}}}^{2} \given
        R_{\tau}}
      &\\ %
      &=& 2\, \E*{\p*{\p*{R_{0}-R_{\tau}} \grad{R_{0}} \payoff_{i}\p{S_{0, R_{0}}}}^{2}
        \given R_{\tau}} %
      &\ + \ && %
      \Order{\E{\abs{S_{0, R_{\tau}} - S_{\tau, R_{\tau}}}^{2} \given R_{\tau}}}  +
      \Order{\tau^{2}},%
    \end{aligned}
  \]} where \(S_{0, R_{\tau}}\) is the solution of~\cref{eq:S-sde} given \(S\p{0}
= R_{\tau}\). Here, both \(\E{\abs{R_{\tau}-R_{0}}^{2}}\) and \(\E{\abs{S_{0, R_{\tau}}
    - S_{\tau, R_{\tau}}}^{2} \given R_{\tau}}\) are \(\Order{\tau}\). Hence, to reduce
the variance of \(\Loss_{i}\), we will \changed{use control variates to
  eliminate the terms involving these factors.
  Starting with the second term, where we use an antithetic variates approach.
  As a general methodology, this is a standard approach to variance reduction
  \cite{glasserman:monte-fin} which has been used previously for pricing
  American options \cite{broadie:mc-america} and also for nested simulation
  within Multilevel Monte Carlo
  \cite{bujok:bernoulli,giles:nested,giles:adaptive}. However, the specific
  treatment used here for estimating portfolio losses does not appear to have
  been previously published.}
    We denote by \(S^{+}\p{\tau}\) and \(S^{-}\p{\tau}\) the two antithetic It\^o
    processes that both start from \(S^{+}\p{0}= S^{-}\p{0} = R_{0}\) and
    depend on the Brownian paths \(\p*{B\p{t}}_{0\leq t \leq \tau}\) and
    \(\p*{-B\p{t}}_{0\leq t \leq \tau}\), respectively.
    Then we set
    \begin{equation}\label{eq:exact-sim-new-loss}
      \widehat \Loss_{i} \defeq  \frac{1}{2} \changed{\p*{
          \payoff_{i}\p{ S_{\tau, S^{+}\p{\tau}}} + \payoff_{i}
          \p{ S_{\tau, S^{-}\p{\tau}}}} - \payoff_{i} \p{S_{\tau, R_{\tau}}}},
    \end{equation}
    where all three processes, \(S_{\tau, R_{\tau}}, \ S_{\tau, S^{+}\p{\tau}}\) and
    \(S_{\tau, S^{-}\p{\tau}}\) use the same Brownian path \(\br*{B\p{t}}_{t\geq\tau}\).
    Then, we have that \(\E*_{\mathbb Q}{\widehat\Loss_{i} \given R_{\tau}} =
    \E*_{\mathbb Q}{\Loss_{i} \given R_{\tau}}\) and, \changed{defining
      \(S^{+}_{t,x}\) to be the solution of~\cref{eq:S-sde} given
      \(S^{+}\p{t} = x\) and using the Brownian path \(\br{B\p{s}}_{s \geq t}\),
      while \(S^{-}_{t,x}\) uses the Brownian path \(\br{-B\p{s}}_{t \leq s \leq
        \tau}\) and \(\br{B\p{s}}_{\max\p{\tau,t} \leq s}\), then for sufficiently
      smooth payoff, \(\payoff_{i}\),
\[
    \begin{aligned}
      \E*{\p*{\widehat \Loss_{i}}^{2} \given R_{\tau}} &= 2\,
      \E*{\p*{\frac{1}{2}\p*{R_{0}-R_{\tau}} \p*{\grad{R_{0}}
            \payoff_{i}\p{S^{+}_{0, R_{0}}} +
            \grad{R_{0}} \payoff_{i}\p{S^{-}_{0, R_{0}}}}}^{2} \given R_{\tau}} \\
      &\hskip1cm + \Order*{\E*{\p*{\frac{1}{2}\p*{S^{+}_{0, R_{\tau}} + S^{-}_{0,
                R_{\tau}}} - S_{\tau, R_{\tau}} \given R_{\tau}}^{2}}} + \Order{\tau^{2}}.
    \end{aligned}
    \]
    Here, assuming the SDE coefficients are sufficiently smooth,} the second
  term is now \(\Order{\tau^{2}} \ll \Order{\tau}\) since \(\tau \ll 1\).

    Finally, similar to \cref{sec:exact-computation}, we can use the Delta
    control variate to eliminate the \changed{remaining} \(\Order{\tau}\) term by defining
    \begin{equation}\label{eq:exact-sim-new-loss-delta}
      \begin{aligned}
        \doublehat{\Loss}_{i} &\defeq \widehat{\Loss}_{i} - \frac{1}{2}\p{R_{0} - R_{\tau}} D_{i}\\
        D_{i} &\defeq \grad{R_{0}} \payoff_{i}\changed{\p{S^{+}_{0, R_{0}}}} +
        \grad{R_{0}} \payoff_{i}\changed{\p{S^{-}_{0, R_{0}}}},
      \end{aligned}
    \end{equation}
    where we assume here that \(\payoff_{i}\p{S}\) is differentiable with
    respect to the initial state, \(R_{0}\). We also modify the loss
    threshold, \(\threshold_{\eta}\), as in~\cref{eq:loss-threshold-delta} so
    that
\[\E*_{\mathbb Q}{\frac{1}{P}\sum_{i=1}^{P}\doublehat{\Loss}_{i} - \widehat \threshold_{\eta}} = \E*_{\mathbb
    Q}{\frac{1}{P}\sum_{i=1}^{P}\Loss_{i} - \threshold_{\eta}},\]
since
\[
  \begin{aligned}
    \grad{R_{0}} V_{i,0} = \E_{\mathbb Q}{ \grad{R_{0}} \payoff_{i}\p{S^{+}_{0, R_{0}}}} = \E_{\mathbb Q}{
    \grad{R_{0}} \payoff_{i}\p{S^{-}_{0, R_{0}}}}.
  \end{aligned}
\]
Recall that \(\grad{R_{0}} V_{i, 0}\) is independent of the risk scenario, \(R_{\tau}\), for all
\(i\) and can be computed once for all risk scenarios as \emph{offline work}.
In summary, to sample \(\doublehat{\Loss}_{i}\), we use all the variance
reduction techniques that were discussed above: (a) the delta control variate
(b) the antithetic pair \(S^{+}\p{\tau}, \) and \(S^{-}\p{\tau}\) and (c) the same
Brownian path \(\br{B\p{t}}_{t \geq \tau}\) when simulating \changed{\(S_{\tau, R_{\tau}}, S_{\tau,
  S^{+}\p{\tau}}\) and \(S_{\tau, S^{-}\p{\tau}}\)}. Indeed, all three variance
reduction techniques %
\changed{ensure that \(\E*{\p*{\doublehat{\Loss}_{i}}^{2} \given R_{\tau}} =\Order{\tau^{2}}\)
compared to \(\E*{\Loss_{i}^{2} \given R_{\tau}} = \Order{\tau}\).}

\subsection{Approximate simulation}\label{sec:nested-approx}
More generally, for some derivatives we might be only able to approximately
sample \(\Loss_{i}\) for a given risk scenario \(R_{\tau}\). This is the case for
example if~\cref{eq:S-sde} cannot be solved analytically and we have to use a
numerical scheme to approximate samples of the \changed{process, \(S\),} and then
compute the loss to obtain an approximate sample of \(\Loss_{i}\).
\changed{The cost per an approximate sample of \(\Loss_{i}\) increases as the
  approximation error, and consequently the bias when estimating \(\E_{\mathbb
    Q}{\Loss_{i} \given R_{\tau}}\), decreases}.

Nevertheless, using Unbiased MLMC~\cite{rhee:unbiased}, we can, in certain
cases, obtain an unbiased \changed{Monte Carlo estimator of
  \changed{\(\E_{\mathbb Q}{\Loss_{i}\given R_{\tau}}\) using samples whose
    expected cost is \(\Order{1}\)}}. To briefly present Unbiased MLMC here,
we denote by \(\Loss_{i, l}\) the \(\nth{l}\) approximation-level of
\(\Loss_{i}\), for example using \(4^{l}\) time steps\footnote{\changed{The
    same discussion applies if \(m^{l}\) time steps are used for the
    \(\nth{l}\) approximation-level, for any \(m>1\). The choice \(m = 4\) is
    motivated by the fact that when the variance of \(\Delta \Loss_{i,l}\)
    decreases like \(m^{-2l}\) while is cost increases like \(m^{l}\), as we
    later assume, this choice minimizes the total cost of an MLMC estimator;
    see \cite{abdullatif:meshMLMC}.}} in a Milstein scheme to approximate the
samples of the solution of~\cref{eq:S-sde}. Then define
\begin{equation}\label{eq:mlmc-delta-def}
  \Delta \Loss_{i, l} \defeq \Loss_{i, l} - \Loss_{i, l-1},
\end{equation}
with \(\Loss_{i, -1} = 0\). As in standard Multilevel Monte Carlo~\cite{giles:acta},
  we assume that the cost of computing \(\Delta \Loss_{i, l}\) grows like \(4^{\gamma l}\) while its
  expectation and variance satisfy, \(\abs{\E_{\mathbb Q}{\Delta \Loss_{i, l}\given R_{\tau}}} =
  \Order{4^{-\alpha l}}\) and \(\changed{\E{\p{\Delta \Loss_{i, l}}^{2}\given R_{\tau}}} = \Order{4^{-\beta l}}\), respectively,
  for \(\alpha, \beta,\gamma> 0\). Then, we write
\begin{equation}\label{eq:umlmc-sum}
  \E_{\mathbb Q}{\Loss_{i} \given R_{\tau}} = \sum_{l=0}^{\infty} \E_{\mathbb Q}{\Delta \Loss_{i, l}\given
    R_{\tau}} = \E*_{\mathbb Q}{ C_{\zeta}\; {4^{\zeta l}} \: \Delta \Loss_{i, l} \given
    R_{\tau}},
\end{equation}
where on the right hand side\changed{, with a slight abuse of notation,} \(l\) is a
random integer satisfying \(\prob{l=j} = 4^{-\zeta j} / C_{\zeta}\) where \({j \in
  \br{0, 1, 2 \ldots}}, \, \zeta>0\) and \( C_{\zeta} \defeq 1/\p{1-4^{-\zeta}} \) is a
normalization constant. In other words, just like the random sub-sampling
method introduced in~\cref{sec:subsampling}, Unbiased MLMC is based on
randomly sub-sampling the corrections \(\Delta \Loss_{i,l}\) to compute the
infinite sum in~\cref{eq:umlmc-sum}. The analysis of Unbiased MLMC is also
similar to the one shown in \cref{sec:subsampling}.
In this setting, the condition \(\gamma < \zeta < \beta \leq 2 \alpha\) is
sufficient~\cite{rhee:unbiased} %
\changed{to bound the expected cost and variance of \(C_{\zeta}\; {4^{\zeta l}} \: \Delta
  \Loss_{i, l}\), for random \(l\) as above, and hence we can estimate
  \(\E_{\mathbb Q}{\Loss_{i} \given R_{\tau}}\) without bias by using standard
  Monte Carlo to estimate \(\E*_{\mathbb Q}{ C_{\zeta}\; {4^{\zeta l}} \: \Delta \Loss_{i,
      l} \given R_{\tau}}\).} %
The optimal value for \(\zeta\), obtained by minimizing the RMS error for a given
computational budget, is \(\p{\beta+\gamma}/2\).
As an example, if \(\payoff_{i}\p{S} \equiv \payoff_{i}\p{S\p{T}}\), for some
maturity \(T>0\), i.e., the payoff is a function of the asset value at
maturity, then if \(\payoff_{i}\) is Lipschitz and a Milstein scheme is used
to approximate samples of the solution of~\cref{eq:S-sde}, then we have
\(\beta=2\alpha=2\gamma\),~\cite{giles:milstein-analysis}. \changed{On the other hand, if
  \(\payoff_{i}\) is discontinuous then one can show that \(\beta = \gamma - \nu\) for
  any \(\nu>0\) using a similar analysis to \cite[Section
  3]{giles:discont-payoff}. In this case, since} \(\beta \leq \gamma\) we would need to
truncate the sum of corrections in~\cref{eq:umlmc-sum} at some maximum level
\(L\) to ensure that Unbiased MLMC has finite work, introducing a bias of
\(\Order{4^{-\alpha L}}\). \changed{A modified Unbiased MLMC estimator
  \cite[Section 4]{rhee:unbiased} can then be constructed with samples which
  have bounded variance but with expected cost that is} \(\Order{ 4^{\p{\gamma-\beta}
    L}}\) for \( \beta < \gamma\) or \(\Order{L^{2}}\) for \(\gamma=\beta\). In the current
work, we will assume that we are always in the case \(\beta > \gamma\). \changed{In the previous
example with a discontinuous \(h_{i}\), an estimator based on conditional
expectation can be used to ensure faster variance convergence~\cite[Section
3.2.8]{giles:milstein-analysis}}.

In summary, in the case of approximate simulation we take \(\Term_{i} \equiv C_{\zeta}
\: 4^{\zeta l}\: \Delta \Loss_{i, l} \) where \(l\) is a random index. In this case,
for a given risk scenario \(R_{\tau}\), the term \(\Term_{i}\) has non-zero
variance and the expected cost to compute it is \(\Order{1}\)\changed{; since we assume
\(\beta > \gamma\)}.

\begin{remark}[Moments of unbiased estimator]\label{rem:unbiased-moments} For the case
    \(\beta > \gamma\), where we do not have to truncate the sum in~\cref{eq:umlmc-sum} and we have an
    unbiased estimator of \(\E_{\mathbb Q}{\Loss_{i} \given R_{\tau}}\), assume further that
    \(\E_{\mathbb Q}{\abs{\Delta \Loss_{i,l}}^{q}} = \Order{4^{-q \beta l / 2}}\) for some \(q > 2 \). The
    \(q\)-moment of the unbiased estimator is then
  \[
    \begin{aligned}
      \E*_{\mathbb Q}{ \abs*{ C_{\zeta}\: 4^{\zeta l} \: \Delta \Loss_{i,l} }^{q} } &=
      C_{\zeta}^{q} \sum_{l=0}^{\infty} {4^{\zeta \p{q-1} l}} \E_{\mathbb Q}{  \abs{\Delta \Loss_{i,l}} ^{q} } \\
      &= \Order*{\sum_{l=0}^{\infty} 4^{{-q \beta l /2 + \zeta \p{q-1}} l}}.
  \end{aligned}
\]
Hence, even if the \(q\)-moment of \(\Delta \Loss_{i,l}\) is finite for a given
level \(l\), the \(q\)-moment of \( 4^{\zeta l} \Delta \Loss_{i,l} \), where \(l\) is a
random level, is finite only when \( q < \p*{1 - {\beta}/\p{2\zeta}}^{-1}\). For
example, when \(\zeta = \p{\beta+\gamma}/2\), the \(q\)-moment of the unbiased estimator is
finite for \(q < 1\!+\!\beta/\gamma\). In other words, if we require certain finite
\(q\)-moments of the unbiased estimator, for example when using MLMC with
adaptive sampling, c.f. \cref{sec:mlmc-adapt}, we might have to use a smaller,
sub-optimal value of~\(\zeta\).
\end{remark}

\paragraph{Control variates}
The discussion on control variates in \cref{sec:exact-sim} carries over to the case of
approximate simulation. Seen another way, we assume we can approximately sample
\(\doublehat{\Loss}_{i}\) in~\cref{eq:exact-sim-new-loss-delta} along with the modified loss
threshold, \(\widehat{\threshold}_{\eta}\), in~\cref{eq:loss-threshold-delta}. Then, denoting the
\(\nth{l}\) approximation-level by \(\doublehat{\Loss}_{i, l}\), and defining \(\Delta
\doublehat{\Loss}_{i, l}\) as in~\cref{eq:mlmc-delta-def}, we set \( \Term_{i} \equiv C_{\zeta}
\:4^{\zeta l}\: \Delta \doublehat{\Loss}_{i,l} \).

One important observation to make here is that, depending on the payoff
function, \(\payoff_{i}\), we might have the case where \(\var{\Delta \doublehat
  \Loss_{i, l} \given R_{\tau}} > \var{\Delta \widehat \Loss_{i, l}\given R_{\tau}}\) for
some \(l\), where \(\Delta \widehat \Loss_{i, l}\) and \(\widehat \Loss_{i,l}\) are
defined as above for \(\widehat \Loss\) in~\cref{eq:exact-sim-new-loss}. In
other words, using the Delta control variate leads to a larger variance for
some approximation levels. As an example, consider \(\payoff_{i}\p{S} =
\payoff_{i}\p*{S\p{T}}\) and \(\payoff_{i}\) is Lipschitz but
\(\grad{R_{0}}\payoff_{i}\) is \changed{discontinuous} and assume that we use the
Milstein scheme to approximate~\cref{eq:S-sde} with \(4^{l}\) time steps.
Then, denote by \(D_{i,l}\) the \(\nth{l}\) approximation-level of \(D_{i}\)
in~\cref{eq:exact-sim-new-loss-delta} and \(\Delta D_{i, l}\) as
in~\cref{eq:mlmc-delta-def} and write
\[
  \Delta \doublehat \Loss_{i, l} \defeq \Delta \widehat \Loss_{i, l} - \frac{1}{2}\p{R_{\tau} -
  R_{0}} \Delta D_{i,l}.
\]
We see that while \(\var{\Delta \widehat \Loss_{i, l} \given R_{\tau}} = \Order{4^{-2
    l}}\), \changed{we have \( \var{\Delta D_{i,l} \given R_{\tau}} =
  \Order{4^{-l\p{1+\nu}}}\) for any \(\nu>0\); using again a similar analysis to
  \cite[Section 3]{giles:discont-payoff}. %
Hence,} for sufficiently large \(l\) we have that \({\var{\Delta \widehat \Loss_{i,
      l} \given R_{\tau}} < \var{\Delta \doublehat \Loss_{i, l} \given R_{\tau}}}\). In
other words, applying the Delta control variate beyond a certain level \(l\)
might lead to an estimator with a larger variance, unless the payoff
\(\payoff_{i}\) is sufficiently smooth; in this example requiring
\(\grad{R_{0}}\payoff_{i}\) to be Lipschitz. An alternative is to use a
modified Milstein scheme for the Delta control variate,~\cite[Section
3.2.8]{giles:milstein-analysis}, so that the variance \(\var{\Delta D_{i,l} \given
  R_{\tau}}\) is sufficiently small compared to, or of the same order as,
\(\var{\Delta \widehat \Loss_{i, l} \given R_{\tau}}\).

If \(\payoff_{i}\) is not sufficiently smooth, then we may apply the Delta
control variate only up to some level, for example, at level \(l=0\) only.
That is, we define
\[
  \widehat{\Delta} \widehat \Loss_{i, l} \defeq
  \begin{cases}
    \doublehat \Loss_{i, l} & l = 0 \\
    \Delta \widehat \Loss_{i, l} & \text{otherwise}
  \end{cases}
\]
and set \(\Term_{i} \equiv \widehat{\Delta} \widehat \Loss_{i, l}\). In this case,
the modification to the threshold value should also be approximated at level
0. That is, we define the new loss threshold
\[
  \doublehat{\threshold\hskip 0.0ex}_{\eta} \defeq \threshold_{\eta} + \frac{1}{2} \, \p{R_{\tau} - R_{0}} \:
  \E_{\mathbb Q}{D_{i,0} \given R_{\tau}},
\]
so that
\[\E*_{\mathbb Q}{\frac{1}{P} \sum_{i=1}^{P} \widehat \Delta \widehat \Loss_{i, l} - \doublehat{\threshold
      \hskip 0.0ex}_{\eta} \given R_{\tau}} =  \E*_{\mathbb Q}{\frac{1}{P} \sum_{i=1}^{P} \Loss_{i} -
    \threshold_{\eta} \given R_{\tau}}.\]

Finally, since the Delta control variate reduces the variance of the first
level only, we should ensure that the variance at level \(l=1\), i.e.,
\(\var{\widehat \Delta \widehat \Loss_{i, 1} \given R_{\tau}}\), is sufficiently
smaller than the variance at level \(l=0\), i.e., \(\var{\doublehat \Loss_{i,
    0}\given R_{\tau}}\), otherwise refining the first level of approximation
of~\cref{eq:S-sde} leads to overall smaller RMS; see the discussion
in~\cite[Section 3]{giles:adaptive} and the end of \cref{sec:mlmc-adapt} for
more details.
\def\mlmcdelta{\Delta \heavisidesymb}
\def\tmlmcdelta{\widetilde \Delta \heavisidesymb}
\section{MLMC and Adaptive Sampling}\label{sec:mlmc-adapt}
The outcomes of the previous section are the terms \( \br*{\Term_{i}}_{i=1}^{P} \)
  and a new loss threshold, \( \doublehat{\threshold\hskip 0.0ex}_{\eta}\), depending on the risk
  scenario, \(R_{\tau}\), such that we can write
\[
  \eta = \prob*{\E*_{\mathbb Q}{\frac{1}{P}\sum_{i=1}^{P} \Loss_{i}  \given R_{\tau}} >
    \threshold_{\eta}} = \E*_{\mathbb P}{ \heaviside*{\E*_{\mathbb Q}{ \frac{\Term_{j}}{P p_{j}} -
        \doublehat{\threshold\hskip 0.0ex}_{\eta} \given R_{\tau}}} },
\]%
where \(j\) is a random integer satisfying \({\prob{j = i}} = p_{i}\) for \(i
\in \br*{1, 2, \ldots, P}\). In this section, for notational convenience, we will
drop the measures \(\mathbb P\) and \(\mathbb Q\), and define the random
variables \(Y \defeq R_{\tau}\) and \( X \defeq {\Term_{j}}/\p{P p_{j}} -
\doublehat{\threshold\hskip 0.0ex}_{\eta} \) so that the objective is to simply
compute \(\E{\heaviside{\E{X \given Y}}}\). Then, we will discuss using MLMC
with adaptive inner sampling as we previously proposed
in~\cite{giles:adaptive}. We start by defining
\begin{equation}\label{eq:inner-MC}
  \inner_{\ell}\p{y} = \frac{1}{N_{\ell}} \sum_{n=1}^{N_{\ell}} X^{(n)}\p{y},
\end{equation}
which is a Monte Carlo estimator of \(\E{X \given Y}\) using \(N_{\ell}\)
samples. Here, \(X^{(n)}\p{y}\) \changed{denotes} the \(\nth{n}\) sample of \(X\)
conditioned on \(Y = y\) and the number of samples \(N_{\ell}\) may depend on
\(y\). Then the MLMC estimator for \({\E{\heaviside{\E{X \given Y}}}}\) is
\[
  \begin{gathered}
    \sum_{\ell=0}^{L} \frac{1}{M_{\ell}} \sum_{m=1}^{M_{\ell}} \mlmcdelta_{\ell} \p{Y^{(\ell, m)}} \\
    \text{where}\quad \mlmcdelta_{\ell}\p{y} = \heaviside{\inner_{\ell}\p{y}} -
    \heaviside{\inner_{\ell-1}\p{y}}
\end{gathered}
\]
and \(\br{Y^{(\ell, m)}}_{\ell, m}\) are i.i.d. samples of \(Y\). Moreover, we set
\changed{\(\heaviside{\inner_{-1}(\cdot)} = 0\)}. We can choose \(N_{\ell}\) uniformly for all \(y\), for
example \(N_{\ell} = N_{0} 2^{\ell}\) for some \(N_{0}>0\). In this case, it can be
shown, \changed{under certain moment and smoothness
  conditions~\cite{giles:adaptive,giorgi:ml2r}}, that
\[
  \begin{aligned}
    \abs{\E{\mlmcdelta_{\ell} \p*{Y}}} &= \Order{2^{-\ell}} \\
    \text{and} \qquad \var{\mlmcdelta_{\ell} \p*{Y}} &= \Order{2^{-\ell/2 }}.
\end{aligned}
\]
Assuming that the expected cost of evaluating \(X\) is \(\Order{1}\)
independently of \(\ell\), the optimal complexity of MLMC to achieve a RMS error,
\(\tol\), can then be shown to be \(\Order{\tol^{-5/2\changed{ - \nu}}}\),~\cite[Theorem
2.1]{giles:acta}.

\changed{To improve the computational complexity, we instead select \(N_{\ell}\)
adaptively based on samples of \(Y\). Let}
\[
  \delta \defeq \frac{\abs*{\E{X \given Y}}}{\p{\var{X \given Y}}^{1/2}}
\]
and let \(\hdelta \approx \delta\) be an estimate computed using Monte Carlo estimates of
\(\E{X \given Y}\) and \(\var{X \given Y}\) for a given \(Y\). \changed{We then select
\(N_{\ell}\) using \cref{alg:adaptive} which is an iterative algorithm that
starts from a minimum number of samples \(N_{\ell} = N_{0} 2^{\ell}\) for a given
\(Y = y\) and then, on every iteration, the number of samples is doubled until
the inequality
\begin{equation}\label{eq:adaptive-N-inequality}
  N_\ell \geq N_0 4^{\ell}\, {\p*{C^{-1} N_0^{\fpow{1}{2}}2^{{\ell}}\hdelta}^{-r}},
\end{equation}
for given constants \(C>0\) and \(1<r<2\), is satisfied or the maximum number
of samples \(N_{0} 4^{\ell}\) is reached. \cref{alg:adaptive}, with
\cref{eq:adaptive-N-inequality}, returns the minimum \(N_{0} 2^{\ell}\) when
\(\hdelta\) is sufficiently large and hence a Monte Carlo estimate of
\(\E{X\given Y}\) is likely to have the correct sign, leading to an exact
evaluation of \(\heaviside{\cdot}\). When \(\hdelta\) is small, estimating the
sign of \(\E{X\given Y}\) using a Monte Carlo estimator is more difficult and
\cref{alg:adaptive} returns a larger number of samples, up to the maximum
\(N_{0} 4^{\ell}\) to account for that; see \cite{giles:adaptive} for a
motivation of the exact form of \cref{eq:adaptive-N-inequality}.}
\begin{algorithm}[ht]
  \caption{Adaptive algorithm to determine $N_\ell$.}\label{alg:adaptive}
  \begin{algorithmic}
    \REQUIRE{$\ell, y, N_0 > 1, C > 0, 1 < r < 2$}%
    \ENSURE{$N_{0} 2^{\ell} \leq N_\ell \leq N_{0}4^{\ell}$}%
    \STATE{set $N_\ell = {N_0 2^{\ell}}$}%
    \STATE{Set \texttt{done}\(\::=\:\)\texttt{false}}%
    \REPEAT%
    \IF{$2 N_\ell \geq N_0 4^{\ell}$}%
    \STATE{Set $N_\ell \equiv N_0 4^{\ell}$}%
    \STATE{Set \texttt{done}\(\::=\:\)\texttt{true}}%
    \ELSE %
    \STATE{Generate $N_\ell$ new, and independent, inner samples of \(X\) given \(Y=y\)} %
    \STATE{Estimate \(\hdelta \approx \delta\)}%
    \IF{\cref{eq:adaptive-N-inequality} is satisfied}%
    \STATE{Set \texttt{done}\(\::=\:\)\texttt{true}}%
    \ELSE%
    \STATE{$N_\ell \equiv 2 N_\ell $}%
    \ENDIF%
    \ENDIF%
    \UNTIL{\texttt{done}}%
    \RETURN $N_{\ell}$%
  \end{algorithmic}
\end{algorithm}
\changed{More concretely}, assuming the following mild conditions:
\begin{itemize}
\item \(\delta\) has a probability density function, \(\rho\), and there exists
  positive constants \(\rho_{0}\) and \(\delta_{0}\) such that \(\rho\p{\delta} \leq \rho_{0}\) for
  all \(\delta \geq \delta_{0}\),
  \item there exists \(q>2\) such that
    \[
    \sup_{y} %
    { \E*{\,
    \p*{\frac{\abs{X \!-\!  \E{X \given Y}\,}}
    {\p*{\var{X \given Y\,}}^{1/2}}}^q \given Y\!=\!y}} <
    \infty
    \]
  \item and \(r\) is chosen such that
    \begin{equation}\label{eq:adaptive-r-bound}
    1 < r < 2 - \frac{\p{4q+1}^{1/2}-1}{q},
  \end{equation}
\end{itemize}
the analysis in~\cite[Theorem 2.7]{giles:adaptive} proves the following two crucial properties
\begin{equation}\label{eq:adaptive-rates}
  \begin{aligned}
    \E{N_{\ell}} &= \Order{2^{\ell}} \\
    \text{and} \qquad \var*{\mlmcdelta_{\ell}\p{Y}} &= \Order{2^{-\ell}}.
  \end{aligned}
\end{equation}
Additionally assuming that the expected cost of evaluating \(X\) is \(\Order{1}\) independently
of \(\ell\) guarantees that the optimal complexity of the MLMC method to achieve a RMS error,
\(\tol\), is \(\Order{\tol^{-2} \abs{\log\tol}^{2}}\), c.f.~\cite{giles:acta, giles:adaptive}.

\paragraph{Antithetic sampling}
Recall that, given a risk scenario \(Y\), we need to sample both
\(\inner_{\ell}\p{Y}\) and \(\inner_{\ell-1}\p{Y}\). Sampling \(\inner_{\ell}\)
requires sampling \(N_{\ell}\) independent and identically distributed samples of
\(X\) given the risk scenario \(Y\). Similarly, sampling \(\inner_{\ell-1}\)
requires sampling \(N_{\ell-1}\) samples of \(X\) \changed{given the same risk
  scenario \(Y\). Here,} \(\var*{\mlmcdelta_{\ell}\p{Y}}\) decreases with
increasing \(\ell\), i.e., with increasing number of internal samples, even if
the internal samples used in \(\inner_{\ell}\) and \(\inner_{\ell-1}\) are mutually
independent. This is because \(\inner_{\ell}\p{Y}\) converges almost surely to
the expectation \(\E{X \given Y}\), due to the Strong Law of Large Numbers.
However, by carefully using the same samples of \(X\) in both \(\inner_{\ell}\)
and \(\inner_{\ell-1}\), we can reduce the variance by a constant factor.

In particular, for a given risk scenario, \(Y\), assume \(N_{\ell} \geq N_{\ell-1}\) and let
\(N_{\ell} = s N_{\ell-1}\) for some integer \(s>0\). Such an integer exists since the
adaptive algorithm always returns \(N_{0} 2^{\hat \ell}\) for some integer \(\hat \ell\). Then, let
\(\br{X^{\p{n}}}_{n=1}^{N_{\ell}}\) be \(N_{\ell}\) samples of \(X\) given \(Y\) and define
\(\inner_{\ell}\p{Y}\) as in~\cref{eq:inner-MC}. Additionally, define \(s\) coarse approximations
as
\[
  \inner_{\ell-1}^{\p{i}}\p{Y} = \frac{1}{N_{\ell-1}} \sum_{n=1}^{N_{\ell-1}} X^{\p{n +
      \p{i-1} N_{\ell-1}}} \p{Y},
\]
for \(i = \br{1, 2, \ldots, s}\). The MLMC estimator with antithetic sampling is
\[
  \begin{gathered}
    \sum_{\ell=0}^{L} \frac{1}{M_{\ell}} \sum_{m=1}^{M_{\ell}} \tmlmcdelta_{\ell} \p{Y^{(\ell, m)}} \\
    \text{where}\quad \tmlmcdelta_{\ell}\p{y} = \heaviside{\inner_{\ell}\p{y}} - \frac{1}{s} \sum_{i=1}^{s}
    \heaviside{\inner_{\ell-1}^{\p{i}}\p{y}}.
\end{gathered}
\]
Note that since \( \E{\tmlmcdelta_{\ell}\p{Y}} = \E{\mlmcdelta_{\ell}\p{Y}} \), the
MLMC estimator with antithetic sampling \changed{has the same expectation}. Moreover,
since \(\tmlmcdelta_{\ell}=0\) whenever \(\inner_{\ell}\) and all
\(\inner_{\ell-1}^{\p{i}}\) for \(i=\br{1, 2, \ldots, s}\) have the same sign, we have
that \(\var{\tmlmcdelta_{\ell}\p{Y}} \leq \var{\mlmcdelta_{\ell}\p{Y}}\). When \(N_{\ell}
\leq N_{\ell-1}\), which may happen due to inaccurate estimates of \(\E{X \given
  Y}\) and \(\var{X \given Y}\), the same discussion as above applies with the
fine approximation having the antithetic estimators instead of the coarse one.

\paragraph{Starting level of MLMC}
An important point to consider when using MLMC is the choice of the starting
level. To explain this, let \(V_{\ell} \defeq \var{\tmlmcdelta_{\ell}\p{Y}}\) and
\(V_{\ell}^{\textnormal{f}} \defeq {\var{\heaviside{\inner_{\ell}\p{Y}}}}\) and let
\(W_{\ell}\) denote the expected work of sampling \( \tmlmcdelta \), in the
current setting we have \(W_{\ell} \equiv \E{N_{\ell}}\). Then, consider the MLMC
estimator
\[
  \frac{1}{M_{0}} \sum_{m=1}^{M_{0}} \heaviside{\inner_{\ell_{0}} \p{Y^{(\ell_{0}, m)}}}
  + \sum_{\ell=\ell_{0}+1}^{L} \frac{1}{M_{\ell}} \sum_{m=1}^{M_{\ell}} \tmlmcdelta_{\ell} \p{Y^{(\ell, m)}}.
\]
In other words, the previous MLMC estimator starts at some level \(\ell_{0} \geq 0\). It can be
shown~\cite{giles:acta} that the \changed{expected} work of MLMC is proportional to
\[
  \p*{\p*{V^{\textnormal{f}}_{\ell_{0}}\, W_{\ell_{0}}}^{1/2} + \sum_{\ell=\ell_{0}+1}^{L}
    \p*{V_{\ell}\, W_{\ell}}^{1/2}}^{2}.
\]
Hence, given some level of approximation, \(L\), an optimal \(\ell_{0}\) satisfies
\begin{equation}
  \p{V^{\textnormal{f}}_{\ell_{0}} \, W_{\ell_{0}}}^{1/2} + \sum_{\ell=\ell_{0}+1}^{\ell_{0}'} \p*{V_{\ell}\, W_{\ell}}^{1/2} %
  < %
  \p*{V^{\textnormal{f}}_{\ell_{0}'} \,W_{\ell_{0}'}}^{1/2},\label{eq:optimal-ell0}
\end{equation}
for all \(\ell_{0} < \ell_{0}' \leq L\). Otherwise, starting at the level \(\ell_{0}'\) leads to overall
less computational work. Since the quantities \(V_{\ell}\) and \(V_{\ell}^{\textnormal{f}}\) for \(\ell
= 0, 1, \ldots, L\) must be approximated using a sample variance estimator, we may relax the
previous condition by multiplying the right hand side by some constant larger than one to
increase the stability of the MLMC algorithm. We use the constant 1.5 in our numerical examples
in \cref{sec:numerics}.

Choosing an optimal starting level is especially relevant in nested simulation applications
because the variance \(V^{\textnormal{f}}_{\ell}\) may be large for small \(\ell\) but then decreases
as more samples are used in the inner estimator, asymptotically converging to
\(\var{\heaviside{\E{X \given Y}}}\). See \cref{sec:numerics} and \cref{fig:portfolio-lvls} for
an illustration of this.
\section{Numerical Experiments}\label{sec:numerics}
In this section, using numerical experiments on fictitious portfolios of put and call options,
we will illustrate the benefits of using random sub-sampling as discussed in
\cref{sec:subsampling}, the control variates that were discussed in \cref{sec:nested}, and
adaptive sampling as discussed in \cref{sec:mlmc-adapt}.

\subsection{Test setup}
\paragraph{Underlying assets}
We assume we have $Q$ assets, $S \equiv \br{S_k}_{k=1}^Q$, modelled by Geometric
Brownian Motions satisfying
\[
  \D S_k(t) = \mu_{k} \, S_k(t)\, \D t + \sigma_{k} \, S_k(t)\,
  \changed{\p[\Big]{\rho\ \D B_{0}(t) + \p{1\!-\!\rho^{2}}^{1/2}\ \D B_i(t)}}
\]
\changed{in the physical measure. Here the Brownian process \(B_{0}\) is the systematic
noise, common to all assets, while \(\br{B_i}_{i=1}^{Q}\) are
mutually independent Brownian processes and represent the idiosyncratic noise
of each asset}. We select the following parameters:
\begin{eqnarray*}
  \mbox{Number of assets:} &\quad&  Q \equiv 16,\\
  \mbox{initial asset price:} & & S_{k}\p{0}   \in  [90, 110],\\
  \mbox{drift rate:}& & \mu_{k} \in [0.05, 0.15],\\
  \mbox{volatility:}& & \sigma_{k} \in [0.01, 0.4], \\
  \mbox{correlation coefficient:}& &  \rho \equiv 0.2 \, .\\
\end{eqnarray*}
\paragraph{Portfolio construction}
The loss of our example portfolio is an average of losses from $P$ derivatives
\cref{eq:total-loss-as-sum}, i.e., \(\Loss \equiv P^{-1} \sum_{i=1}^{P} \Loss_{i} \),
and we consider the market risk. For a short risk horizon, \(\tau = 0.02\), we
set the risk parameter to be the value of the underlying assets at \(\tau\), i.e,
\(R_{\tau} \equiv S\p{\tau}\), and then set
\[
  \Loss_{i} \equiv w_{i} \, \p*{ \payoff_{i}\p{S_{k_{i}}\p{T_{i}}} -
    \payoff_{i}\p{S_{k_{i}, \tau, R_{\tau}}\p{T_{i}}}},
\]
for some weight \(w_{i}\) and \(\payoff_{i}\) being the discounted payoff
function for the \(\nth{i}\) option. Here, \(S_{k, \tau,R_{\tau}}\) is the
\(\nth{k}\) asset conditioned on \(S\p{\tau} = R_{\tau}\). We assume that the
risk-free interest rate is \(r=0.05\) and the discount factor at time \(t\) is
\(\exp\p{-r t}\). Each option is characterized by its type, put or call, which
determines the payoff function \(\payoff_{i}\), along with the following
parameters:
\begin{eqnarray*}
  \mbox{asset:} &\quad &  k_i \in \br{1, 2, \ldots, Q},\\
  \mbox{maturity:} & &  T_i \in [0,5],\\
  \mbox{strike:} & &  K_i \in [80,120],\\
  \mbox{weight:} & &
                    w_i \equiv \begin{cases}
                                   \widetilde w_{i}       & \mbox{put option}\\%
                                   \widetilde w_{i} b_{k_i} & \mbox{call
                                     option.}
                                 \end{cases}
\end{eqnarray*}
To get concrete values for the parameters above, we generate a random instance
of the assets and the portfolio by taking the type to be put or call with
equal probability (ensuring at least a single put and call options for each
underlying asset), and $S_{k}(0), \mu_{k}, \sigma_{k}, k_i, T_i, K_i$ \changed{are
  sampled} independently and uniformly in their respective ranges.
On the other hand, the parameters $b_{k_i}$ are balancing constants which are determined by the
constraint that the portfolio should be delta-neutral with respect to the risk parameter at the
initial time, \(R_{0} = \br{R_{0,k}}_{k=1}^{Q} = \br{S_{k}\p{0}}_{k=1}^{Q}\), i.e.,
\[
  \sum_{i=1}^P \frac{\partial V_{i,0}}{\partial R_{0,k}} = 0, ~~~ \forall k \in \br{1, \ldots, Q}.
\]
More specifically, for $i=1, \ldots, Q$, we set
\[
  b_k \equiv -\frac{\sum_{\substack{i=1 \\ \textnormal{put option}}}^P \frac{\partial}{\partial R_{0,k}}\
    \E{\widetilde w_{i} V_{i,0}}}{\sum_{\substack{i=1 \\
        \textnormal{call option}}}^P \frac{\partial}{\partial R_{0, k}}\ \E{\widetilde w_{i}
  V_{i,0}} }.
\]
We will discuss the choice of \(\br{\widetilde w_{i}}_{i=1}^{P}\) in our
fictitious portfolios below. In any case, the last step is to normalize the
weights, \(\br{w_{i}}_{i=1}^{P}\), so that their average is 1. %

\paragraph{Computation Methods} We consider the three computational models for
computing the value of the options: (a) exact, deterministic evaluation of the
option value using the analytic solution of the Black-Scholes PDE, %
(b) exact simulation of the asset values by analytically solving the SDE, %
and (c) approximate simulation using the Milstein numerical scheme to estimate
the asset values.%

\subsection{Results}
All numerical experiments use MLMC with an initial number of samples of
\(M_{0} = 1024\) to estimate the work and variance of the MLMC levels.
Moreover, for the inner Monte Carlo estimator, we set \(N_{0} = 32\) and, when
using the adaptive algorithm to select the number of inner samples, we set
\(r=1.5\) and \(C=3\) in~\cref{eq:adaptive-N-inequality}. The code was
written in \texttt{C++}\footnote{The full code can be found on
  \url{https://github.com/haji-ali/nested-risk-mlmc}} and the experiments were
carried out in single-precision on an NVIDIA Tesla K20m GPU with 2496
cores\footnote{\changed{Provided by the Edinburgh Centre for Robotics'
    Robotarium Cluster located at Heriot-Watt University, funded by
    Engineering and Physical Sciences Research Council (EPSRC) Centre for
    Doctoral Training in Robotics and Autonomous Systems through grant
    EP/L016834/1.}}. \changed{Note that the embarrassingly parallel nature of
  Monte Carlo simulation makes it possible to fully exploit parallelization in
  addition to the computational savings provided by the sub-sampling
  approach.}

To illustrate the benefit of uniform random sub-sampling we first consider
large, delta-hedged portfolios comprising options with similar nominal values,
i.e., \(\widetilde w_{i} = 1\) for all \( i %
\). The computation method to evaluate each option is chosen to be exact
evaluation or exact simulation with probabilities \(30\%\) and \(70\%\),
respectively.
We compare two methods: (a) in the first method we use random sub-sampling
with uniform probabilities, i.e., setting \(\tg_{i} = 1\) for all \(i\), (b)
and in the second method we do not use any sub-sampling and instead evaluate
the full portfolio for every combination of risk scenarios and underlying
asset values; making sure that options that can be exactly computed are
evaluated only once for every risk scenario. Both methods use MLMC with
adaptive sampling as discussed in \cref{sec:mlmc-adapt}, with appropriate
redefinition of \(X\) and \(Y\), and use all the control variates that were
discussed in \cref{sec:nested}. When estimating the work of these methods, we
simply count the number of times the value of an option or a payoff function
are evaluated; the work estimates are shown in
\cref{fig:large-portfolio-work-est}. For the considered
tolerances, using random sub-sampling leads consistently to fewer evaluations
\changed{and, for a fixed tolerance, the total number of payoff evaluations does not
increase as the number of options increase.}
\cref{fig:large-portfolio-runtime} \changed{shows the actual run-time for the
  numerical tests. Uniform, random} sub-sampling has an overhead that make its
advantage slightly less pronounced for small tolerances \changed{or small
  portfolios}.
To explain these results, recall that evaluating the full portfolio for every
combination of risk scenarios and underlying asset values, i.e, not using
sub-sampling, imposes a minimum budget which increases the computational
complexity for large tolerances. \changed{Nevertheless, for sufficiently small
tolerances or portfolios, and sufficiently large budgets, evaluating the full
portfolio for every risk scenario does not add a significant computational
overhead. On the other hand,} random sub-sampling has an overhead not accounted
for in the work estimate. Namely, the cost of sampling the random option index
which entails sampling a uniform random variable and a table-lookup operation.
While this additional cost is small in typical cases, especially since we use
binary search to perform the table-lookup, it is not wholly insignificant
compared to the cost of sampling the options in our simple numerical example.

\pgfplotstableset{col sep=comma}%

\pgfplotstableread{%
0.00487897342894393,0.00689991039367048,0.00975794685788787,0.013799820787341,0.0195158937157757,0.0275996415746819,0.0390317874315515,0.0551992831493638,0.078063574863103,0.110398566298728,0.156127149726206,0.220797132597455,0.312254299452412,0.441594265194911,0.624508598904824,0.883188530389822,
0.00182639511736789,0.000543219401646857,0.000305755604134601,0.00291516838668655,0.00203999038301698,0.00387431994457021,0.00372936593150475,0.00398830997036431,0.0229760555149259,0.00855931403597262,0.0581567610542559,0.0716418065073952,0.116397917991482,0.160692987705986,0.185920450370415,0.199839492921515,
0.00344995519683524,0.00487897342894393,0.00689991039367048,0.00975794685788787,0.013799820787341,0.0195158937157757,0.0275996415746819,0.0390317874315515,0.0551992831493638,0.078063574863103,0.110398566298728,0.156127149726206,0.220797132597455,0.312254299452412,0.441594265194911,0.624508598904824,0.883188530389822
0.00167468813366849,0.00170209228212183,0.00175014113208005,0.00167021236405983,0.0039840307038999,0.000858334771961372,0.00111830173688984,0.00751194866455251,0.0339076677099804,0.0464304608510359,0.040316253563628,0.0175029157760393,0.0563879794922738,0.0346282959188927,0.0534297035161412,0.104213815766644,0.172200804871202
0.00267986664485167,0.00382848765112418,0.00492365965151407,0.00698086217029676,0.0099578746217467,0.0155379847251186,0.0206947934050097,0.0321886014753152,0.0428377135418804,0.0613689229577618,0.103944682403641,0.150980611584201,0.191427341215286,0.247823184008437,0.328688910172187,0.370296301713589,
0.00320264394473825,0.00409971305544749,0.00522126596551383,0.00607068853025938,0.00951685221556311,0.0164876347624487,0.0190935915477902,0.0246349546096498,0.0503428088836157,0.0640720406385573,0.0877011757194408,0.115714294673897,0.205476706309649,0.229709483212565,0.330572879898366,0.410065642439976,0.718807498812237
1711584606.89134,1717505865.90938,1720954331.94345,1729219021.88952,1727624313.03626,1701113845.39325,1758704924.89089,1781578124.90275,1802590759.20453,1643716430.72788,1663968238.90288,1771203223.17375,2168603459.00112,3043304507.1714,4839186163.25785,9065727462.81749,
143752547.348049,143497483.300413,143763355.693672,141754528.737788,119731793.126141,104385959.838316,103584309.213631,102117255.183567,73761464.6117346,73844235.4741319,70816385.9324654,69292001.4568885,45300338.1166454,50445119.7312668,46759684.3595122,54658299.2543054,35314802.4945859
3.6881956741974,3.69977759240424,3.70654998473611,3.72353166362911,3.71840696537329,3.6644977083876,3.78731073955877,3.83892753380772,3.8882709638359,3.55530249035963,3.61466705003755,3.85309623553976,4.7198467025972,6.63403449395628,10.5545388920417,19.559488410556,
0.64779360391484,0.647236362462845,0.649191496609215,0.641850841576824,0.549843704980411,0.484556642799215,0.484522314033808,0.481373556533471,0.356681208181701,0.363692273472322,0.35886440662335,0.363688202434873,0.250119787024079,0.286282234436354,0.272652273198271,0.326194883190878,0.195362560778838
0,1,2,3,4,5,6,7,8,9,10,11,12,13,,,
0.0289398339073606,6.88991439825301e-06,,,,,,,,,,,,,,,
9.83278725103162e-05,4.55158340177001e-05,,,,,,,,,,,,,,,
0.229789642486558,0.0282567049808429,0.0878310086921366,0.0566236403280344,0.0197442401733636,0.0055927168088757,0.00148199871278098,0.000387832134670554,8.56748026158634e-05,3.09543206502138e-05,,,,,,,
0.0122579470062583,0.0102134463160492,0.0105139219718305,2.23385289399372e-05,2.80972905394271e-05,2.92452909880269e-05,2.94187379847085e-05,2.92448929474786e-05,2.90818304329418e-05,2.90766413934137e-05,,,,,,,
0.028102319920775,0.000317389298690708,,,,,,,,,,,,,,,
9.40917615344095e-05,2.16357502043042e-05,,,,,,,,,,,,,,,
0.176986362692458,0.102854082851708,0.0607615578579476,0.0326565916508612,0.0147927420355605,0.00671810069818573,0.00318813588583459,0.0015601465945104,0.000770658182855233,0.000389397949934088,,,,,,,
0.00839504642683142,0.00592429822908934,0.00548298418212129,1.02981316508974e-05,1.28238487433208e-05,1.36015484824914e-05,1.38304778669308e-05,1.3814450321573e-05,1.37336733362687e-05,1.38638182731095e-05,,,,,,,
0.028102319920775,0.0280797545673535,,,,,,,,,,,,,,,
9.40917615344095e-05,0.000409688740532706,,,,,,,,,,,,,,,
0.176986362692458,0.162097171992296,0.0999161967880352,0.0522943070656908,0.0347970327776223,0.0298271163549325,0.0285559769983165,0.028229257290236,0.0281513896121861,0.028031157483777,,,,,,,
0.00839504642683142,0.009189833239961,0.0112822376306484,2.57792432128492e-05,4.0626098677464e-05,5.87221603601251e-05,8.41577055768553e-05,0.000119000679820468,0.00016817552099245,0.000236094424108289,,,,,,,
2270000,9080000,,,,,,,,,,,,,,,
32,128,620.827167980594,2797.18049459892,9648.20620810404,24927.8964221637,53849.7616518342,107808.807622535,212450.111361856,420596.372290886,,,,,,,
0.00489102619600389,0.0195748291426524,,,,,,,,,,,,,,,
1.08449979274186e-06,1.83695685038675e-06,4.31139059010867e-06,1.34319758918145e-05,4.40546166845203e-05,0.000112569769080378,0.000242355399814825,0.000527716965911822,0.000905589870704513,0.00168928581661166,,,,,,,
0.606517073719786,,,,,,,,,,,,,,,,
1.31902732115781,1.35817204587756,1.44144075729564,1.31208312229429,1.1465528521533,1.02949075139473,0.945915775226505,0.878798483408693,0.830100651977265,,,,,,,,
1,1,1,1,1,1,1,1,1,1,1,1,1,1,1,1,
6,6,6,6,6,5,5,5,4,4,4,4,3,3,3,3,2
0,0,0,0,0,0,0,0,0,0,0,0,0,0,0,0,
3,3,3,3,3,3,3,3,3,3,3,3,3,3,3,3,3
0.0074753461928961,0.0105717359694277,0.0149506923857922,0.0211434719388555,0.0299013847715844,0.042286943877711,0.0598027695431688,0.084573887755422,0.119605539086338,0.169147775510844,0.239211078172675,0.338295551021688,0.478422156345351,0.676591102043376,,,
0.00377850763487808,0.00463827059162589,0.00336193628070004,0.00151302369356802,0.00341835764823085,0.0092685607764749,0.00808532225330844,0.00873497709801772,0.0531960181665458,0.0941593078889306,0.11247958785501,0.0719769812469476,0.00462724348812486,0.0942204309892149,,,
0.0211434719388555,0.0299013847715844,0.042286943877711,0.0598027695431688,0.084573887755422,0.119605539086338,0.169147775510844,0.239211078172675,0.338295551021688,0.478422156345351,0.676591102043376,,,,,,
0.0017269739768019,6.45266079073656e-05,0.00360528401877105,0.00488174781640808,0.0208741356179445,0.0311887421287273,0.0235578811304802,0.072232445400796,0.00430531162353447,0.135946383563037,0.038220300215659,,,,,,
0.0149506923857922,0.0211434719388555,0.0299013847715844,0.042286943877711,0.0598027695431688,0.084573887755422,0.119605539086338,0.169147775510844,0.239211078172675,0.338295551021688,0.478422156345351,0.676591102043376,,,,,
0.00165048137263307,0.00278123555750848,1.46675698000313e-05,0.00410246986027703,0.0160356831166419,0.0171734038799165,0.0211655463419275,0.055589273701197,0.000826897120858098,0.0150585995121315,0.00546847761605917,0.0828193974505902,,,,,
0.00485791202742864,0.0041357517763177,0.0043673447750604,0.0041185213909018,0.00222582017356161,0.0179604673124505,0.0136661684112203,0.00796375598925284,0.019806386774808,0.0028464767275142,0.0121565901487791,0.0461108820332664,0.0846807160344834,0.00301628890641714,,,
0.00155869883331047,0.00149638467413282,0.00184697576134691,0.00583736213337614,0.0138930995064628,0.0264623671491919,0.044867657112896,0.0207049157402726,0.0177453995635345,0.0524558349484525,0.0102940868968391,0.131951603373883,0.0880967052566945,0.308797764096764,,,
0.00620180591223235,0.00781270541068572,0.0112739336331047,0.0166146117995354,0.0274282134891732,0.0391364636875882,0.0467861514492109,0.0524196372173148,0.111019157630626,0.129809097629439,0.175131969970134,0.335636941012021,0.439074391809544,0.487369576550637,,,
0.0170693530242163,0.0270555483316247,0.0325032226939194,0.0423305464137031,0.0836505454106143,0.0961803795993748,0.139721319915351,0.186029020011334,0.322009613767837,0.406834298396856,0.505265251509699,,,,,,
0.0121965373477403,0.0142116588595904,0.0297088245454178,0.0384138050384406,0.0497759409419747,0.0646418913814257,0.0697179319632794,0.111117115793511,0.192664113259221,0.265675810632814,0.37983779536474,0.471554671468731,,,,,
0.00504025242265908,0.0103213260778053,0.0129447535452715,0.0164265661476578,0.0195532921875261,0.0374445080974093,0.0495271929865758,0.0644914491063314,0.0876622206498515,0.165819596544579,0.217673586231538,0.283693700689505,0.452846664690663,0.555364460103765,,,
0.00568108528172848,0.00714634918174896,0.00932671613536202,0.0176757177322864,0.0209176597230712,0.0266736157493583,0.037068214272587,0.0779162545418663,0.0974983232596698,0.118535906258129,0.239185226426947,0.302746933263759,0.365247580299547,0.612999725364299,,,
7784371827.71375,7805242970.89808,7608475583.52497,6096863332.26242,3504748331.82156,3382875656.20343,3386707728.03586,5035346255.1081,1559039098.81337,1623989842.27533,1601687077.56717,557121918.630254,489477667.958801,549239214.621418,,,
42271729791.4141,37167081482.72,36243355444.8202,30986088832.0299,23786931771.5873,24017173374.9014,25691214590.6596,22062009414.7776,13207404940.1074,11983667644.0943,15636220924.2383,,,,,,
45111736015.3616,56763809263.839,30320607235.5194,29099671235.6733,29135470097.5943,27361517373.3117,26250743975.981,23979358117.6567,19048697973.8314,18940153527.2191,19241613813.3126,23420627303.0172,,,,,
1151367425.56529,937367123.627711,937441687.134301,950532426.549545,1060312175.36532,633147275.491701,633394336.939502,644901033.665101,755525930.391382,389674459.995349,396275744.121565,342644727.214274,260964214.838931,231244812.871504,,,
21388670086.3813,21323748869.0112,22996068930.2792,14935437379.0051,14797489347.5102,14571111466.3991,14442050454.4131,10227975484.1393,10170515097.6485,10116526840.9658,5307310209.58256,7307358519.20487,6976341125.38326,4268317473.30548,,,
31.4074933339856,31.4896953935224,30.5852339148711,24.6327884476864,14.5467911476102,14.0906083297629,14.1362569833247,20.6932256712158,6.75125883158564,7.02953520870466,6.95504983266528,2.64899609627429,2.34724787628213,2.59546913787255,,,
165.163560284816,144.875046655821,141.962318878308,122.604897821953,93.3527552473412,94.7858349300706,102.158149102718,89.1249887664611,52.697955372515,48.2469968807924,63.588875131176,,,,,,
75.3074906031598,93.6205317262686,54.3356952276249,53.6416586817378,55.7542484009306,55.5062567429567,57.8193529897839,58.5841225427171,52.7685406820413,57.5037047241685,63.976473860227,78.6839336693876,,,,,
5.37386945150174,4.43254586744041,4.44985183016299,4.51042231995925,5.04100014016991,3.12732358776483,3.16540182830689,3.26025539552192,3.82231819828133,2.13412507139391,2.24336539956018,2.08028469073266,1.75368943287684,1.64229405808431,,,
40.1119603690754,40.010334794875,43.2325985623887,28.3367288361178,28.1813448583245,27.7955202391776,27.5803363595141,19.7797898353932,19.7059292932691,19.6104071968929,10.5248738970008,15.5727753996196,15.2214235827747,9.97447146199769,,,
0.31924387473633,0.0544303797468354,0.107810961601575,0.0869186046511628,0.0380589914367269,0.0155171893820373,0.00466827004629137,0.00126033691179075,0.000315133048546774,6.57498680400148e-05,2.32100621145472e-05,,,,,,
0.0125970031663519,0.0101471022668749,0.0149993173451766,0.0115040718201303,0.0145834661664976,1.63138028596698e-05,2.20660060746228e-05,2.60130320369658e-05,3.08703454630757e-05,3.66808627614129e-05,4.38837358969781e-05,,,,,,
0.489531313823409,0.0198288741002309,0.0368932782116062,0.0570088468578401,0.105977082128398,0.119224767848988,0.0853719420868697,0.04437255859375,0.0120245314222712,0.00404705122204399,0.00102555199545226,0.000273306505267144,8.51047404647827e-05,6.08548894875207e-05,,,
0.00544371257094994,0.00680032970366428,0.00629934092294669,0.0104309236155751,0.00998716919248706,0.0105793330815992,0.0143076541731806,0.0138421659771528,0.00633290155346354,6.97199236578518e-05,9.52954244391194e-05,9.40298432609043e-05,9.35105284248796e-05,0.000147898282240315,,,
0.136467889908257,0.0728293664096547,0.037841796875,0.01446533203125,0.00571379435920571,0.00165199695868742,0.000380946056947507,0.000126100654906622,8.55810549063152e-06,,,,,,,,
0.0201352427007997,0.0117548665608428,0.0141934607445617,0.00969274547070186,4.86933032082474e-05,6.43729547480515e-05,6.42861901266996e-05,6.46807806165422e-05,6.42831721683853e-05,,,,,,,,
0.313080881788156,0.0540061388653683,0.097431294020163,0.0804037352375721,0.0326757448789572,0.017102878501343,0.00569474340084614,0.00174594273061094,0.000488957847948776,0.000128406526489871,2.83044837877404e-05,4.39550534196289e-06,,,,,
0.0086218114177436,0.010382984153588,0.0100133673215999,0.0109294961264035,0.0139481276807093,0.00165868189000101,2.35807541091705e-05,3.07821488355923e-05,3.05391708155066e-05,3.02963615765099e-05,3.00497773700072e-05,3.65315625716774e-05,,,,,
0.250412870423633,0.0206633774607982,0.0180662983425414,0.0658175192519252,0.0775980777877698,0.0556315316786414,0.0262013976363367,0.00984488588238171,0.00313430992846286,0.000844798893408382,0.000228447598479803,5.16114373073734e-05,1.56376709814009e-05,,,,
0.00559889750548773,0.00724695833344426,0.0103138475755399,0.0097455200978846,0.010461458672192,0.0142708270339516,0.00144293864822748,2.30963804331592e-05,2.93408987642042e-05,2.98299377164396e-05,3.03309082602626e-05,3.09985551178075e-05,3.16672219675933e-05,,,,
0.217327223179664,0.112974042621375,0.0761680666602831,0.0505846910492158,0.0248360041318087,0.010519088602156,0.0047757732327008,0.00230025367338678,0.00114115902797926,0.000568436161094801,0.000290401547988125,,,,,,
0.00743135995985025,0.00550674768988099,0.00777161708124677,0.00577004964957221,0.00671183520656595,7.50207334388603e-06,1.0357480930338e-05,1.22990496208932e-05,1.46454003318753e-05,1.74772550903602e-05,2.08856465393842e-05,,,,,,
0.249890406609736,0.151332505009014,0.143978420550438,0.13870113482542,0.114971063547212,0.0816908347597079,0.0455591180869422,0.0218004547059536,0.00808351497152796,0.00410677271493563,0.00195248707819719,0.000949454477928288,0.000487689379016781,0.000224652263707189,,,
0.00272364548216469,0.0036107322580391,0.00328367586464018,0.005486978705111,0.00531825259149295,0.00550439280239903,0.00661098161650465,0.0059381592237393,0.0028638687577283,3.24109171338724e-05,4.49355915212606e-05,4.48114046129783e-05,4.45423544644078e-05,7.22318733865427e-05,,,
0.117844404932245,0.0457979618328872,0.0229210257530212,0.0106893442571163,0.00520321869197367,0.00243892733191272,0.00116025860478154,0.000565559028548365,0.000276635824183556,,,,,,,,
0.0161893710470585,0.00580234418683468,0.00663127708590857,0.00448228480076604,2.24780973609034e-05,3.00932182936697e-05,3.02696009370287e-05,3.05234722006046e-05,3.01928098031518e-05,,,,,,,,
0.215061243246907,0.113172720961468,0.0806707943749171,0.0483257330429005,0.0232163317184861,0.011153468631929,0.00528141553841754,0.00252792674294568,0.00123319569324844,0.000611070922220187,0.000301690872294393,0.000149700600334351,,,,,
0.00513570752061449,0.00558288475423843,0.00525247941682333,0.00526052707740533,0.00635408849236293,0.000742317761610958,1.08916249708865e-05,1.43597462260573e-05,1.43200707726631e-05,1.43345360423446e-05,1.41515173175188e-05,1.71531074038108e-05,,,,,
0.18770626474983,0.119456030058819,0.106966426543756,0.0767399308784443,0.0540887265402574,0.0346442342200739,0.0156109127425751,0.00715966652576048,0.00330311865495282,0.00157343555703093,0.000769469232387075,0.000388718907009558,0.000194833407638266,,,,
0.00370070271179485,0.00423138354097648,0.00572263506413459,0.0052090274053521,0.00516490956519924,0.00664345794350346,0.000645311994950973,1.04759748596854e-05,1.35125389938251e-05,1.38933754574074e-05,1.41928689225678e-05,1.45665207890639e-05,1.48259518445294e-05,,,,
0.217327223179664,0.19237569363884,0.126106165856456,0.0678893996755003,0.0418522163206443,0.019370073607913,0.014886374859299,0.0136296398324956,0.0133784057074154,0.0132033747681319,0.0131135270229701,,,,,,
0.00743135995985025,0.00861058659650948,0.0152172845334485,0.0118929624391932,0.0177029224686133,2.14848255119075e-05,3.80781236940936e-05,6.20125732938865e-05,0.000103556082859562,0.000173246693650203,0.000289033784594382,,,,,,
0.249890406609736,0.249017859051637,0.245638864817268,0.235952996606109,0.196211088948848,0.133306824952057,0.0662508795468804,0.033099889755249,0.0173293834916228,0.0145154221853666,0.0136337241026118,0.0132414125084728,0.0131200432704664,0.0129192000573173,,,
0.00272364548216469,0.00438726004792628,0.00422028401617208,0.00734213259074701,0.0083542500716734,0.0104430231299789,0.0153791286589083,0.0161528685250242,0.00902819752978586,0.000128029363699737,0.000246522355023461,0.000344066969542893,0.000475326263476072,0.00109961840833713,,,
0.117844404932245,0.0680261115367235,0.0339202880859375,0.0158376693725586,0.0151112705234334,0.0136233472010493,0.0132696343274626,0.013314418357595,0.0133135157642552,,,,,,,,
0.0161893710470585,0.0127810818105897,0.0163642847900359,0.0113741889046329,8.0828265213417e-05,0.000148885993678193,0.000212982831949742,0.000307485224339011,0.000436953361620675,,,,,,,,
0.215061243246907,0.192730768479331,0.138166516793411,0.0751946479718721,0.0410723760182267,0.0204812567804303,0.0153242761490814,0.0137942323883652,0.0133507908977538,0.0132649561863585,0.0133127448567856,0.0132725609287716,,,,,
0.00513570752061449,0.00880002868634475,0.0100273691157263,0.0119392021542181,0.0173417081437837,0.00216018575519525,3.9044867775187e-05,7.02496832734566e-05,9.83424782301148e-05,0.000138244481765517,0.000195531832937154,0.000337012685184483,,,,,
0.18770626474983,0.177749312183258,0.188627465584079,0.146092017841086,0.0904152994909603,0.0540915036555642,0.0257800701205317,0.0170158674247111,0.0142139298017234,0.0134713562500556,0.0132651639029405,0.0133201026047395,0.0132245164555523,,,,
0.00370070271179485,0.0060394697910188,0.00902408397821108,0.0100225793486865,0.0114040124268671,0.0160262119316821,0.00175830913107232,3.44386187106881e-05,5.93803710013975e-05,8.53976140458134e-05,0.000123330971041967,0.000177747166197565,0.000255635331868138,,,,
32,128,512,2048,8192,32768,131072,524288,2097152,8388608,33554432,,,,,,
32,128,637.695176848875,2931.4153229321,12472.6338924234,50910.2901507079,198179.019470794,679888,1933835.85446527,4537018.76798123,9417933.37046912,18392024.8830798,35404528.0791465,68035276.0123333,,,
22700,90800,357054.00390625,1147396.09375,3002110.86932879,6740548.69103681,14128412.8691504,29069709.3902004,59346720.3749716,,,,,,,,
32,128,626.370943239884,2682.79703378193,9449.11731843575,27512.9485282026,65701.4101903137,137446.285831232,272243.059503252,531223.031887716,1051208.10039236,2083898.9861424,,,,,
68.5947560817205,301.179229153437,1410.33016574586,6549.56655665567,25179.7913669065,86056.5486610059,262208.392649674,686115.708766872,1628208.75525358,3650615.45657835,7963872.02018356,17078526.0711599,36272386.3777529,,,,
1.07947947935259e-06,2.08381459682803e-06,4.53433483438019e-06,1.27691862195037e-05,4.85790898752258e-05,0.000174225766393526,0.000693836837471189,0.00199903170040113,0.00773119012235498,0.031960818448967,0.127336149931321,,,,,,
7.08985284379033e-07,1.5117463802621e-06,4.11790612097667e-06,1.58619945096874e-05,6.37481500814249e-05,0.000258990629094252,0.00106083912309027,0.00272964150644839,0.00767660311853478,0.0175458863626287,0.0364143479508477,0.0718619963998644,0.139901317863146,0.280216597569789,,,
5.0820979867871e-05,0.000201545143702907,0.0007506487891078,0.00271975924260914,0.0054097869182264,0.0123863931491053,0.0268578679161421,0.0394044790532555,0.0807459900275222,,,,,,,,
9.92012844353798e-07,1.93561163118186e-06,4.7584278710951e-06,1.69150156818946e-05,6.56718870337005e-05,0.00017133195861706,0.000352761656844866,0.000669350005258223,0.00119426887228987,0.00224491146060301,0.00445981539584817,0.00878257798432799,,,,,
9.36287646723573e-07,1.92813790366865e-06,4.73783819714962e-06,1.69028018829715e-05,6.55156352537141e-05,0.000243511175034795,0.000662427328010247,0.00138020486614651,0.00305225452382292,0.0067661227163161,0.0144910057159354,0.0310536008508873,0.0662651672132861,,,,
1.29776441429492,1.39473092512952,1.54464930339256,1.40715131002325,1.47188504639735,1.13675501456252,0.933357725083021,0.796732141900169,0.710793796044345,0.650522559347189,,,,,,,
1.28043758534819,1.2166888921681,1.24258999450981,1.29710895102769,1.38331584695371,1.54822144262308,1.57538083119021,1.50244627608841,1.24525439813942,1.09459936667776,0.993886116149859,0.916875894227906,0.858830182916287,,,,
1.47860467605865,1.53617032559033,1.63792607745041,1.21969875648158,1.1259830006888,0.995561226300631,0.90207667661302,0.844048883279016,,,,,,,,,
1.29446608991688,1.29801570695751,1.45665294099607,1.47280328190554,1.56784554086297,1.33518066820376,1.15681070838214,1.02616529917672,0.932823196319346,0.860137737142823,0.817518223019246,,,,,,
1.31020542254776,1.20163976146755,1.25204788664242,1.42174504543925,1.49964006122088,1.6079995838266,1.4095844112034,1.19231819650367,1.02775700540688,0.923137901614833,0.852288549149624,0.810032416344534,,,,,
5,5,5,5,4,4,4,4,3,3,3,2,2,2,,,
4,3,3,3,2,2,2,2,1,1,1,,,,,,
4,4,3,3,3,3,3,3,2,2,2,2,,,,,
5,4,4,4,4,3,3,3,3,2,2,2,2,2,,,
5,5,5,4,4,4,4,3,3,3,2,3,3,2,,,
5,5,5,5,5,5,5,5,5,5,5,5,5,5,,,
9,9,9,9,9,9,9,9,9,9,9,,,,,,
4,4,4,4,4,4,4,4,4,4,4,4,,,,,
6,6,6,6,6,6,6,6,6,6,6,6,5,5,,,
7,7,7,7,7,7,7,7,7,7,7,6,6,6,,,
}\Oloadedtable%

\pgfplotstabletranspose{\loadedtable}{\Oloadedtable}
\pgfplotstableset{col sep=space}%

\begin{figure*}[htb]
  \centering
  \begin{tabular}{lcl}
    \ref{P-conv-work/full-sampling-ref} {No sub-sampling}
    & \ref{P-conv-work/subsampling-ref} {Uniform sub-sampling}
  \end{tabular}
  \vskip .2cm
  \begin{subfigure}[t]{0.49\textwidth}
    \axisonleft%
\begin{tikzpicture}

\begin{axis}[
xlabel={\tol},
ylabel={Work estimate $\times \tol^2$},
xmin=0.002, xmax=1.5,
ymin=1e7, ymax=2e10,
xmode=log,ymode=log
]
\addplot [mark=diamond]
table [x=0, y=6] {\loadedtable};
\label{large-portfolio/total_work-line0}

\addplot [mark=square]
table [x=2, y=7] {\loadedtable};
\label{large-portfolio/total_work-line1}

\coordinate (legend) at (axis description cs:0.91,0.5);
\end{axis}

\end{tikzpicture}
    \phantomsubcaption\label{fig:large-portfolio-work-est}
  \end{subfigure}
  \begin{subfigure}[t]{0.49\textwidth}
    \axisonright%
\begin{tikzpicture}

\begin{axis}[
xlabel={\tol},
ylabel={Runtime, [s] $\times \tol^2$},
xmin=0.002, xmax=1.5,
ymin=0.1, ymax=38,
xmode=log,ymode=log,
]
\addplot [mark=diamond]
table [x=0, y=8] {\loadedtable};
\label{large-portfolio/total_time-line0}

\addplot [mark=square]
table [x=2, y=9] {\loadedtable};
\label{large-portfolio/total_time-line1}

\coordinate (legend) at (axis description cs:0.03,0.97);
\end{axis}

\end{tikzpicture}
    \phantomsubcaption\label{fig:large-portfolio-runtime}
  \end{subfigure}

  \begin{subfigure}[t]{0.49\textwidth}
    \axisonleft%
\begin{tikzpicture}

\begin{axis}
[
xlabel={\(P\)},
ylabel={Work estimate $\times \tol^2$},
xmode=log,
ymode=log]
\addplot [mark=square]
table {%
1000 94501827.2175221
2000 150616716.540416
4000 130115047.744692
8000 132742466.780639
16000 123937849.002678
32000 186745952.148095
64000 167274200.592532
128000 163590221.328631
};
\label{P-conv-work/subsampling-ref}
\addplot [mark=diamond]
table {%
1000 114132676.212018
2000 338360151.502718
4000 258543175.511742
8000 432937234.528314
16000 507264121.99793
32000 700037685.552917
64000 1117014302.22197
128000 2085643256.87754
};
\label{P-conv-work/full-sampling-ref}
\end{axis}

\end{tikzpicture}
    \phantomsubcaption\label{fig:portfolios-P-work-est}
  \end{subfigure}
  \begin{subfigure}[t]{0.49\textwidth}
    \axisonright%
\begin{tikzpicture}

\begin{axis}
[
xlabel={\(P\)},
ylabel={Runtime, [s] $\times \tol^2$},
xmode=log,
ymode=log
]
\addplot [mark=square]
table {%
1000 0.356027679335119
2000 0.579321417315063
4000 0.510949178043547
8000 0.526040358125375
16000 0.494907538357582
32000 0.757456624604541
64000 0.715375940495823
128000 0.712227708312014
};
\label{P-conv-time/subsampling-ref}
\addplot [mark=diamond]
table {%
1000 0.214253728603011
2000 0.632844841385195
4000 0.492356845675641
8000 0.828880013980233
16000 0.991338585600876
32000 1.38796904220304
64000 2.25600904070095
128000 4.25349380829965
};
\label{P-conv-time/full-sampling-ref}
\end{axis}

\end{tikzpicture}
    \phantomsubcaption\label{fig:portfolio-P-runtime}
  \end{subfigure}\vskip -0.4cm

  \caption{ The work estimate (\emph{left}), measured in number of evaluations of option values
    and payoff functions, and runtime (\emph{right}), measured in seconds, of MLMC with
    adaptive sampling when applied to large portfolios of options with similar
    nominal values, i.e., \(\widetilde w_i =1\) for all \(i\). 30\% of the options are computed
    using exact evaluation while \(70\%\) are computed using exact simulation. Here \(\tol\) is
    the tolerance normalized by the exact value which was estimated using
    Monte Carlo to be 3-4\%
    approximately for the considered portfolios.
    Note that the work estimates and
    running time are multiplied by \(\tol^{2}\) to \changed{normalize the work effort
    for different portfolios and to emphasize the difference of the
    computational effort when using random, uniform sub-sampling or not.
    In the (\emph{top}) plots we fix the size of portfolio to \(P=10^{5}\) and
    vary \(\tol\), while in the (\emph{bottom}) plot we fix \(\tol \approx 3 \times 10^{-3}\) and vary
    \(P\).
    We see that using random sub-sampling, even when applied to options with
    similar nominal value, reduces the computational complexity, particularly
    for large tolerances. Moreover the computational complexity is independent
    of the number of options in the portfolio.}
  }\label{fig:large-portfolio-total-works}
\end{figure*}

Random sub-sampling is most useful when the financial derivatives in the
portfolio are heterogeneous, even in moderate-sized portfolios. To illustrate
this we consider a smaller portfolio of \(10^{3}\) options with different
nominal values. To model this, we sample the logarithm of the weight
parameters, \(\log\p{\widetilde w_{i}}\), from a normal distribution with mean
0 and standard deviation 3. Moreover, when using random sub-sampling we use
the estimates \(\tg_{i} = \widetilde w_{i}\). Like before, the computation
method of each portfolio is chosen to be exact evaluation or exact simulation
with probabilities \(30\%\) and \(70\%\), respectively. We now test several
methods and show their work estimates and runtimes in
\cref{fig:portfolio-total-works}.

The first method, labelled ``Full method'', uses MLMC with adaptive sampling
as discussed in \cref{sec:mlmc-adapt}, all the control variates as discussed
in \cref{sec:nested} and random sub-sampling as discussed
in~\cref{sec:subsampling}. The second method, labelled ``No sub-sampling''
does not use random sub-sampling and instead evaluates the whole portfolio for
every combination of risk scenarios and asset values; again making sure that
options that can be exactly computed are evaluated once for every risk
scenario. In this case, the work reduction measured by work estimates and
total runtime is more than tenfold. The third method we consider, labelled
``No CV'', is the same as ``Full method'' except that we do not use the Delta
and antithetic control variates that were discussed in \cref{sec:nested}. In
this example, by using these control variates, work estimate and runtime is
again reduced by around 40-fold. Recall that this reduction is related to the
risk horizon, \(\tau=0.02\), and we should expect that longer risk horizons,
compared to the maturities of options, would reduce the savings of the
antithetic and Delta control variates. The fourth method we consider, labelled
``Non-adaptive'', is again the same as ``Full method'' except that it uses
instead deterministic, non-adaptive number of inner samples, i.e. \(N_{\ell} =
N_{0} 4^{\ell}\) for all risk scenarios. Using adaptive sampling is two to seven
times more efficient than non-adaptive sampling. Moreover, recall that to
achieve RMS error \(\tol\), we expect MLMC with adaptive sampling to have a
computational complexity of \(\Order{\tol^{-2} \abs{\log\tol^{-1}}^{2}}\)
while MLMC with non-adaptive sampling would have a complexity of
\(\Order{\tol^{-5/2}}\)\changed{, approximately}. The observed complexities in
\cref{fig:portfolio-total-works} are consistent with the expected complexities
and with the variance and work estimates in \cref{fig:portfolio-lvls}.

To show that using the framework outlined above accommodates approximate
simulation, we also include in these plots the runtime of the ``Full method''
when applied to a similar portfolio with the same number of options and the
same weights but with the computational method being exact evaluation, exact
simulation or approximate simulation with probabilities \(30\%\), \(50\%\) and
\(20\%\), respectively. Recalling the discussion in
\cref{rem:unbiased-moments} and the notation used there, we note that setting
\(r=1.5\) in the adaptive algorithm to select the number of inner samples
would not work in this setting. This is because we use the Milstein scheme to
approximate samples of the underlying assets for \(20\%\) of the options,
which yields \(\beta = 2\gamma\), and we use Unbiased MLMC with \(\zeta = \p{\beta+\gamma}/2\) to
approximate the expectation of the loss, as discussed in
\cref{sec:nested-approx}. Hence, the \(q\)-moments of the unbiased estimator
are finite for \(q<3\) only while \(r=1.5\) requires finite \(q\)-moments for
\(q\geq 15\) to satisfy the condition~\cref{eq:adaptive-r-bound}. Instead, we
set \(r=1.1\) in this case which requires finite \(q\)-moment for \(q \approx
2.72\).

The starting levels, \(\ell_{0}\), of MLMC for each of the methods in this
section were selected based on the criteria~\cref{eq:optimal-ell0}. As
discussed above, a correct choice of the starting level is crucial in nested
simulation because the variance, \(V_{\ell}^{\textnormal{f}} =
\var{\heaviside{\inner_{\ell}\p{X \given Y}}} \) may exhibit a pre-asymptotic
behaviour with respect to \(\ell\). This is illustrated in
\cref{fig:portfolio-lvls}-\emph{(top)}.

\newcommand\portfoliolegend[1]{
  \begin{tabular}{lll}
    \ref{portfolio/total_work-line0} {Non-adaptive} &%
                                                      \ref{portfolio/total_work-line1} {No CV} &%
                                                                                                 \ref{portfolio/total_work-line2} {No subsampling} \\%
    \ref{portfolio/total_work-line3} {Full method} & %
                                                     \multicolumn{2}{l}{\ref{portfolio/total_work-line4} {Full method with approximate simulation}}#1
    \end{tabular}
    \vskip 0.1cm
}
\begin{figure*}[htb]
  \centering %
  \portfoliolegend{ \\ \ref{portfolio/total_work-ref-adapt} \(\Order{\tol^{-2}
      \abs{\log\tol}^{2}}\) & \ref{portfolio/total_work-ref-nonadapt} \(\Order{\tol^{-5/2}}\) &}
  \begin{subfigure}[t]{0.49\textwidth}
    \axisonleft%
\begin{tikzpicture}

\begin{axis}[
xlabel={\tol},
ylabel={Work estimate $\times \tol^2$},
xmin=0.005, xmax=1,
ymin=175619211.575015, ymax=74743169231.7063,
xmode=log,ymode=log
]
\addplot [clr_non_adaptive, mark=diamond]
table [x=33, y=46] {\loadedtable};
\label{portfolio/total_work-line0}

\addplot [clr_no_cv, mark=square]
table [x=35, y=47] {\loadedtable};
\label{portfolio/total_work-line1}

\addplot [clr_no_subsampling, mark=*]
table [x=37, y=48] {\loadedtable};
\label{portfolio/total_work-line2}

\addplot [clr_full, mark=triangle]
table [x=33, y=49] {\loadedtable};
\label{portfolio/total_work-line3}

\addplot [clr_approx, mark=asterisk]
table [x=33, y=50] {\loadedtable};
\label{portfolio/total_work-line4}

\addplot [dashed, domain=0.0001:1] {%
  3e7 * (ln(x*0.2))^2};
\label{portfolio/total_work-ref-adapt}

\addplot [dotted, domain=0.0001:1] {%
  8e8 * x^(-0.5)};
\label{portfolio/total_work-ref-nonadapt}

\coordinate (legend) at (axis description cs:0.09,0.5);
\end{axis}

\end{tikzpicture}
    \phantomsubcaption\label{fig:portfolio-work-est}
  \end{subfigure}
  \hfill
  \begin{subfigure}[t]{0.49\textwidth}
    \axisonright%
\begin{tikzpicture}

\begin{axis}[
xlabel={\tol},
ylabel={Runtime, [s] $\times \tol^2$},
xmin=0.005, xmax=1,
ymin=1.3041506310692, ymax=207.987579966451,
xmode=log,ymode=log
]
\addplot [clr_non_adaptive, mark=diamond]
table [x=33, y=51] {\loadedtable};
\label{portfolio/total_time-line0}

\addplot [clr_no_cv, mark=square]
table [x=35, y=52] {\loadedtable};
\label{portfolio/total_time-line1}

\addplot [clr_no_subsampling, mark=*]
table [x=37, y=53] {\loadedtable};
\label{portfolio/total_time-line2}

\addplot [clr_full ,thick, mark=triangle]
table [x=33, y=54] {\loadedtable};
\label{portfolio/total_time-line3}

\addplot [clr_approx, mark=asterisk]
table [x=33, y=55] {\loadedtable};
\label{portfolio/total_time-line4}

\coordinate (legend) at (axis description cs:0.03,0.97);

\addplot [dashed, domain=0.0001:1] {%
  0.15 * (ln(x*0.2))^2};
\label{portfolio/total_time-ref-adapt}

\addplot [dotted, domain=0.0001:1] {%
  3 * x^(-0.5)};
\label{portfolio/total_time-ref-nonadapt}

\end{axis}

\end{tikzpicture}
    \phantomsubcaption\label{fig:portfolio-runtime}
  \end{subfigure}
  \caption{ The work estimate (\emph{left}) and runtime (\emph{right}) of MLMC with adaptive
    sampling when applied to a portfolio of \(10^{3}\) heterogeneous options. Here \(\tol\) is
    the tolerance, normalized by the exact value which was estimated using Monte Carlo to be
    1\% approximately for our particular portfolio. Note that the work estimates and running
    time are multiplied by \(\tol^{2}\) to emphasize the differences between the methods, since
    \(\Order{\tol^{-2}}\) is the computational complexity in the best-case when the inner
    expectation can be computed exactly at \(\Order{1}\) cost. The full method, which uses MLMC
    with adaptive inner sampling, all control variates as discussed in \cref{sec:nested} and
    random sub-sampling with non-uniform probabilities, clearly outperforms other
    the methods.}\label{fig:portfolio-total-works}
\end{figure*}
\begin{figure*}[htb]
  \centering
  \portfoliolegend{}
  \begin{subfigure}[t]{0.49\textwidth}
    \axisonleft%
\begin{tikzpicture}

\begin{axis}[
xlabel={$\ell$},
ylabel={$V_\ell^{\textnormal{f}}$},
xmin=-1, xmax=14,
ymin=0.01, ymax=0.310114784343749,
ymode=log,ylabel shift=-3ex
]
\addplot [mark=diamond, error bars/.cd, y dir=both, y explicit]
table [x=10, y=76] {\loadedtable};
\label{portfolio/fine_var-line0}

\addplot [mark=square, error bars/.cd, y dir=both, y explicit]
table [x=10, y=78] {\loadedtable};
\label{portfolio/fine_var-line1}

\addplot [mark=*, error bars/.cd, y dir=both, y explicit]
table [x=10, y=80] {\loadedtable};
\label{portfolio/fine_var-line2}

\addplot [mark=triangle, error bars/.cd, y dir=both, y explicit]
table [x=10, y=82] {\loadedtable};
\label{portfolio/fine_var-line3}

\addplot [mark=asterisk, error bars/.cd, y dir=both, y explicit]
table [x=10, y=84] {\loadedtable};
\label{portfolio/fine_var-line4}

\coordinate (legend) at (axis description cs:0.97,0.97);
\end{axis}

\end{tikzpicture}
     \phantomsubcaption\label{fig:portfolio-fine-var}
  \end{subfigure}
  \hfill
  \begin{subfigure}[t]{0.49\textwidth}
    \axisonright%
\begin{tikzpicture}

\begin{axis}[
xlabel={$\ell$},
ylabel={$V_\ell$},
xmin=-1, xmax=14,
ymin=9.08588463711389e-05, ymax=0.368520629758317,
ymode=log
]
\addplot [clr_non_adaptive, mark=diamond, error bars/.cd, y dir=both, y explicit]
table [x=10, y=66] {\loadedtable};
\label{portfolio/var-line0}

\addplot [clr_no_cv, mark=square, error bars/.cd, y dir=both, y explicit]
table [x=10, y=68] {\loadedtable};
\label{portfolio/var-line1}

\addplot [clr_no_subsampling, mark=*, error bars/.cd, y dir=both, y explicit]
table [x=10, y=70] {\loadedtable};
\label{portfolio/var-line2}

\addplot [clr_full, mark=triangle, error bars/.cd, y dir=both, y explicit]
table [x=10, y=72] {\loadedtable};
\label{portfolio/var-line3}

\addplot [clr_approx, mark=asterisk, error bars/.cd, y dir=both, y explicit]
table [x=10, y=74] {\loadedtable};
\label{portfolio/var-line4}

\negratetriangle{1}{1e-3}{3}{1}{\(\Order{2^{-\ell}}\)}

\coordinate (legend) at (axis description cs:0.97,0.97);

\end{axis}

\end{tikzpicture}
     \phantomsubcaption\label{fig:portfolio-var}
  \end{subfigure}

  \begin{subfigure}[t]{0.49\textwidth}
    \axisonleft%
\begin{tikzpicture}

\begin{axis}[
xlabel={$\ell$},
ylabel={Work},
xmin=-1, xmax=14,
ymin=15.4443903499008, ymax=140965669.933915,
ymode=log
]
\addplot [clr_non_adaptive, mark=diamond]
table [x=10, y=86] {\loadedtable};
\label{portfolio/work-line0}

\addplot [clr_no_cv, mark=square]
table [x=10, y=87] {\loadedtable};
\label{portfolio/work-line1}

\addplot [clr_no_subsampling, mark=*]
table [x=10, y=88] {\loadedtable};
\label{portfolio/work-line2}

\addplot [clr_full, mark=triangle]
table [x=10, y=89] {\loadedtable};
\label{portfolio/work-line3}

\addplot [clr_approx, mark=asterisk]
table [x=10, y=90] {\loadedtable};
\label{portfolio/work-line4}

\coordinate (legend) at (axis description cs:0.97,0.03);

\posratetriangle{12}{1e5}{3}{1}{\(\Order{2^{\ell}}\)}
\posratetriangle{6}{1e2}{3}{2}{\(\Order{4^{\ell}}\)}

\end{axis}

\end{tikzpicture}
     \phantomsubcaption\label{fig:portfolio-work}
  \end{subfigure}
  \hfill
  \begin{subfigure}[t]{0.49\textwidth}
    \axisonright%
\begin{tikzpicture}

\begin{axis}[
xlabel={$\ell$},
ylabel={Runtime, [s]},
xmin=-1, xmax=14,
ymin=3.7221544436854e-07, ymax=0.53374852419889,
ymode=log,
]
\addplot [mark=diamond]
table [x=10, y=91] {\loadedtable};
\label{portfolio/time-line0}

\addplot [mark=square]
table [x=10, y=92] {\loadedtable};
\label{portfolio/time-line1}

\addplot [mark=*]
table [x=10, y=93] {\loadedtable};
\label{portfolio/time-line2}

\addplot [mark=triangle]
table [x=10, y=94] {\loadedtable};
\label{portfolio/time-line3}

\addplot [mark=asterisk]
table [x=10, y=95] {\loadedtable};
\label{portfolio/time-line4}

\coordinate (legend) at (axis description cs:0.97,0.03);

\posratetriangle{11}{2e-4}{3}{1}{\(\Order{2^{\ell}}\)}
\posratetriangle{6}{1e-6}{3}{2}{\(\Order{4^{\ell}}\)}

\end{axis}

\end{tikzpicture}
     \phantomsubcaption\label{fig:portfolio-time}
  \end{subfigure}
  \caption{ \emph{(top)} The variance estimates of the MLMC levels where \(V_{\ell} \defeq
    \var{\tmlmcdelta\p{Y}}\) and \(V_{\ell}^{\textnormal{f}} \defeq
    \var{\heaviside{\inner_{\ell}\p{Y}}}\). Note that \(V_{\ell}^{\textnormal{f}}\) has a
    pre-asymptotic behaviour where it asymptotically approaches \(\var{\heaviside{\E{X \given
          Y}}}\) from above. Because of this, the starting level should be chosen carefully as
    discussed in \cref{sec:mlmc-adapt}. Note also that \(V_{\ell}\) decreases like
    \(\Order{2^{-\ell}}\) for all methods.\\
    \emph{(bottom)} Work estimate and runtime of the MLMC levels. Note that the work increases
    like \(\Order{2^{\ell}}\) for methods that use adaptive inner sampling for sufficiently large
    \(\ell\), unlike the non-adaptive method where the work increases like \(4^{\ell}\) for all
    \(\ell\). Additionally, when not using the control variates and because of the increase of the
    variance per level, the region of pre-asymptotic behaviour where the work increases like
    \(4^{\ell}\) is extended. }\label{fig:portfolio-lvls}
\end{figure*}

\section{Conclusions}
This work has shown the application of MLMC with adaptive sampling to estimating the
probability of a large loss of a large financial portfolio of heterogeneous derivatives. The
key elements to reduced computational complexity are using MLMC with adaptive sampling,
applying several control variates that exploit the short risk horizon and using sub-sampling
strategies to obtain a computational complexity that does not depend on the number of
derivatives in the portfolio.
Using the methods above to efficiently compute probabilities of loss in a
portfolio, other risk measures such as Value-at-Risk (VaR) or Conditional VaR
(CVaR) can also be computed efficiently as discussed in detail
in~\cite{giles:adaptive}. VaR can be computed by finding the root
\(\threshold_{\eta}\) of the equation \(\prob{\E{\Loss \given R_{\tau}} >
  \threshold_{\eta}} = \eta \) for a given risk level, \(\eta\). Given an efficient
method to solve the forward problem, i.e., computing \(\eta\) given an estimate
of \(\threshold_{\eta}\), the root can be approximated efficiently using a
stochastic root finding algorithm, c.f.~\cite{giles:adaptive}. \changed{Since CVaR can
be written as a minimization problem whose solution is VaR
\cite{giles:adaptive,rockafellar:cvar}, then we can write, denoting \(X \defeq
\E{\Loss \given R_{\tau}}\),
\[
  \begin{aligned}
    \E*{X \given X > \threshold_{\eta}}
    &= \threshold_\eta + \eta^{-1} \E*{ \max(0,X\!-\!\threshold_\eta)} \\
    &= \min_x \br{x + \eta^{-1} \E*{ \max(0,X \!-\! x)}} \\
    &= \widetilde{\threshold}_\eta + \eta^{-1} \E*{\max(0,X \!-\!
      \widetilde{\threshold}_\eta)} +
    \Order{\widetilde{\threshold}_\eta\!-\!\threshold_\eta}^2,
\end{aligned}
\]}%
given an estimate of VaR, \(\widetilde{\threshold}_\eta \). Hence, to
approximate CVaR, we first approximate \(\widetilde{\threshold}_\eta\) up to a
RMS error \(\tol^{1/2}\) with work \(\order{\tol^{-2}}\). Then,
\changed{\(\E*{\max(0,\E{\Loss \given R_{\tau}} \!-\!
    \widetilde{\threshold}_\eta)}\), involving a nested expectation, can be
estimated with total work \(\Order{\tol^{-2}}\) to achieve a RMS error
\(\tol\) using MLMC with antithetic sampling for nested
expectations~\cite{bujok:bernoulli,giles:nested,giles:adaptive} combined with random
sub-sampling of the financial derivatives in the portfolio and the control
variates that were discussed in \cref{sec:subsampling,sec:nested},
respectively.}

\bibliographystyle{siam}

\end{document}